\documentclass[
amsmath,amssymb,
aps,
prab,
reprint, 
nofootinbib, 
preprintnumbers,
]{revtex4-2}

\usepackage{graphicx}	
\usepackage{dcolumn}	
\usepackage{siunitx}	
\usepackage{bm}			

\DeclareMathOperator{\sinci}{si}
\DeclareMathOperator{\erf}{erf}

\begin{document}


\title{Pickup concepts for ultra-low charged short bunches in X-Ray Free-Electron Lasers}


\author{B.~E.~J.~\surname{Scheible}}
\email[]{bernhard.scheible@iem.thm.de}
\altaffiliation[Also at ]{Technische Universität Darmstadt, Schloßgartenstr. 8, 64289 Darmstadt, Germany.}
\affiliation{Technische Hochschule Mittelhessen, Wilhelm-Leuschner-Straße 13, 61169 Friedberg, Germany.}
\author{M.~K.~\surname{Czwalinna}}
\author{H.~\surname{Schlarb}}
\affiliation{Deutsches Elektronen-Synchrotron DESY, Notkestrasse 85, 22607 Hamburg, Germany.}
\author{W.~\surname{Ackermann}}
\author{H.~\surname{De Gersem}}
\affiliation{Technische Universität Darmstadt, Schloßgartenstr. 8, 64289 Darmstadt, Germany.}
\author{S.~\surname{Mattiello}}
\author{A.~\surname{Penirschke}}
\affiliation{Technische Hochschule Mittelhessen, Wilhelm-Leuschner-Straße 13, 61169 Friedberg, Germany.}
\date{\today}

\begin{abstract}
The all-optical synchronization systems used in various X-ray free-electron lasers (XFEL) such as the European XFEL observe the transient fields of passing electron bunches coupled into one or more pickups in the Bunch Arrival Time Monitors (BAM). The extracted signal is then amplitude modulated on reference laser pulses in a Mach-Zehnder type electro-optical modulator. With the emerging demand for future experiments with ultra-short FEL shots, fs precision is required for the synchronization systems even with \SI{1}{\pico\coulomb} bunches. Since the sensitivity of the BAM depends in particular on the slope of the bipolar signal at the zero-crossing and thus, also on the bunch charge, a redesign with the aim of a significant increase by optimized geometry and bandwidth is inevitable. In this contribution the theoretical foundations of the pickup signal are aggregated and treated with a focus on ultra-short bunches as well as a general formulation. A possible new pickup concept is simulated and its performance is compared to the previous concept. A significant improvement of slope and voltage is found. The improvement is mainly achieved by the reduced distance to the beam and a higher bandwidth.
\end{abstract}

\maketitle

\section{Introduction}
Free-electron lasers (FEL) became an important light source for experiments in various fields since they provide ultra-short pulses with extreme brilliance in atomic length and time scales \cite{jaeschke-2020}. FEL are well suited for applications in pump-probe experiments \cite{jaeschke-2020}, where the timing jitter is specifically critical \cite{Seddon-2017}, as well as for capturing image sequences with atomic resolution on \si{\femto\second}-time-scales, even below the FEL repetition rate \cite{jaeschke-2020, Gunther-2011, Lu-2018}.

For the generation of ultra-short X-ray pulses, FEL with short and ultra-low charge electron bunches (\(\leq\)\SI{1}{\pico\coulomb}) have been found as a favorable option \cite{Rosenzweig-2008, Reiche-2008}. Short bunches may shorten the X-ray pulse, reduce timing jitter and lead to single-spike operation, if sufficiently small compared to the cooperation length of the SASE process \cite{Seddon-2017, Rosenzweig-2008, Reiche-2008}. The {European XFEL} (EuXFEL) was upgraded from initially \SI{1}{\nano \coulomb} electron bunches to cover a range from \SIrange{0.02}{1}{\nano \coulomb} \cite{Decking-2013} with a possible bunch length below \SI{3}{\femto \second} in the undulator section \cite{Tschentscher-2011}. Moreover, a decrease to ultra-low charges of \SI{1}{\pico \coulomb} is targeted.

The application in time-resolved experiments entails tight requirements for the overall machine synchronization in order to reduce the timing jitter \cite{Seddon-2017}. The timing information is also used for post-processing experimental data \cite{jaeschke-2020}. The synchronization concerns all critical subsystems, specifically in the injector and if present the seeding and the pump laser \cite{jaeschke-2020}. Furthermore, the instrumentation must be suited for a broad spectrum of operation modes with different bunch properties even in a single bunch train \cite{Kot-2013, Viti-2017}. Besides, bunch arrival time monitors (BAM) are installed throughout the whole facility, thus experiencing different bunch properties.

A tremendous improvement in synchronization, exceeding RF techniques, and reduction of arrival time as well as energy jitter was achieved by the implementation of an all-optical synchronization system with two different feedback loops \cite{Lohl-2010}. Though some updates have been introduced \cite{Angelovski-2012, Angelovski-2013}, the basic scheme in use by the \textit{Deutsches Elektronen-Synchrotron} (DESY) remained unchanged.

In this contribution the all-optical synchronization system will be introduced with special attention on the state-of-the-art cone-shaped pickups. The schematic BAM description is followed by an in-depth analytical discussion of the voltage signal, regarding different limiting cases as well as a general solution, to identify principal parameters determining the BAM resolution and their effect on the signal shape. For design purposes an additional numerical treatment is introduced. These considerations are the basis for three designs, which are presented at the end of this paper.

\section{All-optical Synchronization System}
The all-optical synchronization system, as successfully tested at the {Free-Electron Laser in Hamburg} (FLASH) by L\"ohl \cite{Lohl-2010}, mainly comprises of a mode-locked reference laser, length stabilized fiber links and different end-stations for synchronization and arrival time measurement \cite{Kim-2006, Lohl-2010, Angelovski-2012, Angelovski-2013, Viti-2017, Lamb-2018}. The arrival time is non-destructively measured with respect to the reference laser in the BAM, which include high-bandwidth pickup electrodes in the RF unit, a Mach-Zehnder type electro-optical modulator (EOM) and the data acquisition system (DAQ) \cite{Viti-2017, Lamb-2018}.

\subsection{Basic working principle of BAM}
The transient electric fields of passing electron bunches are extracted in the RF unit and, if foreseen, initially processed with analogue components like RF combiners, limiters or attenuators \cite{Lohl-2010, Angelovski-2012}. The received bipolar signal is transmitted over radiation hard silicon dioxide coaxial cables to the EOM \cite{Angelovski-2012}, there it is probed at its zero-crossing by the reference laser \cite{Lohl-2010}. Any temporal deviation will lead to an additional voltage, which the EOM turns into an amplitude modulation of the laser pulse. Therefore, the laser amplitude holds the timing information, which can be retrieved in the DAQ \cite{Lohl-2010}. The signal slope at its zero-crossing strongly influences the BAM's temporal resolution. The minimum design requirement for the currently installed BAM was \(\geq \SI{300}{\milli\volt\per\pico\second}\) with \SI{20}{\pico\coulomb} bunches \cite{Angelovski-2012}.

\subsection{RF unit and pickups}
The RF unit so far comprises four identical pickups mounted circularly around the beam line. The combination of opposite pickup signals compensates for the orbit dependency \cite{Angelovski-2015}. The original pickups used at FLASH were of button-type and designed for a \SI{10}{\giga\hertz} bandwidth \cite{Kuhl-2011}. This pickup struggled with ringing and strong signal reflections at the alumina vacuum feedthrough \cite{Kuhl-2011} degrading the signal strength and resolution for charges below \SI{150}{\pico\coulomb} \cite{Bock-2011}. Hence, a new design was required \cite{Bock-2011}.

\begin{figure}
	\includegraphics[width=\columnwidth]{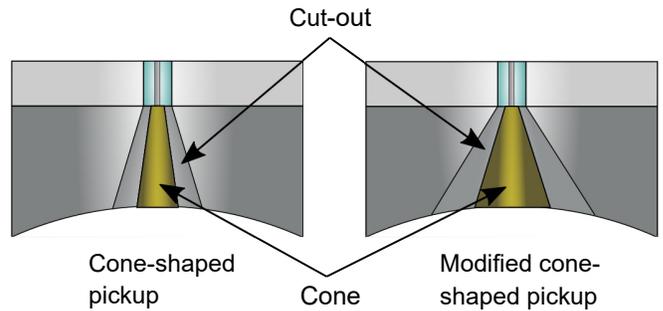}
	\caption{Cross section of the 1\textsuperscript{st} generation cone-shaped pickup (left) and the modified 2\textsuperscript{nd} generation cone-shaped pickup for high peak-to-peak voltage (right) adapted from \cite{Angelovski-2013}.\label{fig:UpdatedPickup}}
\end{figure}

A novel pickup (Fig. \ref{fig:UpdatedPickup}, left), similar to those designed for the CERN linear collider test facility \cite{Yin-1995}, was proposed as a solution in FEL applications \cite{Angelovski-2011}. The cone-shaped design, finalized in \cite{Angelovski-2012}, with \SI{40}{\giga\hertz} bandwidth became a new standard device used at the EuXFEL and at other FELs.

Due to high losses in the RF path a design update was necessary (Fig. \ref{fig:UpdatedPickup}, right). The second generation of cone shaped pickups were optimized towards a maximum signal voltage at the cost of its slope by increasing the active surface and letting the cone slightly protrude into the beam pipe \cite{Angelovski-2013}. A limit was reached when a larger surface, e.g. more than a few bunch lengths, further decreases the slope without a further increase in peak-to-peak voltage \cite{Angelovski-2013}. The protrusion increases the inductance, which unfavorably deforms the signal shape \cite{Angelovski-2013}. Nonetheless, a combination of both modifications increased the peak-to-peak voltage sufficiently for \SI{20}{\pico\coulomb} bunches while maintaining an acceptable slope \cite{Angelovski-2013}.

Recently the performance of the state-of-the-art system was evaluated at the EuXFEL. The correlation of two adjacent monitors with less than \SI{1}{\meter} distance was analyzed. Examining the measured arrivaltimes for a period of \SI{1}{\minute} gave a timing jitter of approximately \SI{6}{\femto\second} r.m.s. caused by the BAM resolution and critical parts of the reference laser distribution system \cite{Schulz-2019}.

\section{RF signal}

The limitations of current pickup structures are evident in the theoretical consideration of the time domain voltage signal. 
The image charge on the pickup surface is calculated by
\begin{equation}\label{eq:PUimageCharges}
	Q_\mathrm{im}\left(t\right) = \int_{-\infty}^{\infty} \frac{\lambda(z-c_\mathrm{0}t)w(z)}{2 \pi r_\mathrm{p}} \mathrm{d}z ,
\end{equation}
with the line charge density \(\lambda(t)\) of the ultra-relativistic bunch (\(v \approx c_0\)), pickup width \(w(z)\) and the distance between pickup surface and bunch \(r_\mathrm{p}\) \cite{Smith-1996}.
Because line charge density and pickup width are real quantities, by substitution of \(z\) into a moving frame, Eq. \eqref{eq:PUimageCharges} can be transformed into the cross-correlation 
\begin{equation}\label{eq:PUimageCharges-Corr}
	Q_\mathrm{im}\left(t\right) = \left(\lambda \star w \right)(t) , 
\end{equation}
which is readily solved numerically. This is true for any bunch and pickup form. In the special case of an even charge density, Eq. \eqref{eq:PUimageCharges} is the convolution
\begin{equation}\label{eq:PUimageCharges-Conv}
	Q_\mathrm{im}\left(t\right) = \left(\lambda^\mathrm{even} \circledast w \right)(t) . 
\end{equation}
Eqs. \eqref{eq:PUimageCharges-Corr} and \eqref{eq:PUimageCharges-Conv} can be utilized in the numerical treatment and also for analytical solutions, which are accessible for many different cases. Some essential examples are covered in this section followed by the numerical approach.

\subsection{Analytical Model}
Idealized representations of bunch and pickup surface, allow to find an analytical solution for Eq. \eqref{eq:PUimageCharges}. A suitable representation that leads to a substantial simplification is given by a Gaussian bunch and a rectangular pickup surface. The stationary charge density of the Gaussian beam centralized on the \(z\)-axis when is
\begin{equation}
	\rho(x,y,z) = \delta(x) \delta(y) \lambda(z) .
\end{equation}
The line charge density is found by integration of the transverse area. A Gaussian bunch is given by
\begin{subequations}
\begin{equation}\label{eq:GaussianBunchProfile}
	\lambda(z) = \lambda_0 \exp \left(- \frac{z^2}{2 \sigma_{\mathrm{z}}^2} \right) = Q_{\mathrm{b}} \ G(z,\sigma_{\mathrm{z}}),
\end{equation}
where \(\lambda_0 = Q_{\mathrm{b}}/\sqrt{2 \pi \sigma_{\mathrm{z}}^2}\) is a constant for normalization by the bunch charge, \(Q_{\mathrm{b}}\) the total charge of one bunch, \(\sigma_{\mathrm{z}}\) the r.m.s. length and \(G(\tau)\) the normalized Gaussian distribution function
\begin{equation}
	G(\tau,\sigma_{\tau}) = \frac{1}{\sqrt{2 \pi} \sigma_{\tau}} \exp\left(-\frac{\tau^2}{2 \sigma_{\tau}^2}\right).  
\end{equation}
In the scope of this work, we denote 
\begin{equation}
		\lambda(t) := \left.\lambda(z)\right|_{z = c_0 t} = \frac{Q_{\mathrm{b}}}{c_0} \ G(t,\sigma_{\mathrm{t}}),
\end{equation}
\end{subequations}
using Eq. \eqref{eq:GaussianBunchProfile}, where the speed of light is incorporated into the r.m.s length in units of time
\begin{equation}\label{Eq:Def-Sigt}
	\sigma_{\mathrm{t}} = \frac{\sigma_{\mathrm{z}}}{c_0} .
\end{equation}
The rectangular pickup surface width is written as
\begin{equation}\label{eq:RectangularPickup}
	w(z) = w_0 \ \Pi_{\ell} \left( z \right) ,
\end{equation}
where \(\ell\) is the longitudinal extension of the pickup and 
\begin{equation}
	\Pi_{\ell} (z) = 
	\begin{cases}
		1 & \left|z\right| \leq \frac{1}{2} \ell \\
		0 & \left|z\right| > \frac{1}{2} \ell
	\end{cases}
\end{equation}
the rectangular function.

\subsubsection{Long Bunch Approximation}
In many applications, specifically with hadrons, it is reasonable to assume a bunch much longer than the pickup. This is treated in the long bunch approximation (LBA). The pickup form is insignificant in the LBA, hence a more general approach is used. For this purpose the parameterization of a finite pickup profile is
\begin{equation}\label{eq:IndicatorPickup}
	w\left(z\right) = \tilde{w}\left(z\right) \chi_{\left[a,b\right]} (z), 
\end{equation}
where \(\chi_{\left[a,b\right]}\left(z\right)\) is the indicator function of the interval \({\left[a,b\right]}\), which bounds the pickup in longitudinal direction, and \(\tilde{w}\left(z\right)\) defines the pickup width at location \(z\). The rectangular pickup in Eq. \eqref{eq:RectangularPickup} is a special case of Eq. \eqref{eq:IndicatorPickup}. The longitudinal characteristic dimension of the pickup is defined by \(\ell = \left| a - b \right|\) and its surface area \(A_{\mathrm{p}}\) by the integral
\begin{equation}\label{Eq:PickupSurface-General}
	A_{\mathrm{p}}  = \int_{-\infty}^{\infty} \tilde{w}\left(z\right) \chi_{\left[a,b\right]} (z) \mathrm{d}z = \int_{a}^{b} \tilde{w}\left(z\right) \mathrm{d}z. 
\end{equation}
The image charge found by introducing Eq. \eqref{eq:IndicatorPickup} in Eq. \eqref{eq:PUimageCharges} is
\begin{subequations}
\begin{equation}
	Q_\mathrm{im}\left(t\right) = \frac{1}{2 \pi r_\mathrm{p}}\int_{-\infty}^{\infty}  \lambda(z-c_\mathrm{0}t) \tilde{w}\left(z\right) \chi_{\left[a,b\right]} (z) \mathrm{d}z	
\end{equation}
which becomes
\begin{equation}
	Q_\mathrm{im}\left(t\right) = \frac{\lambda_{\mathrm{c}}}{2 \pi r_\mathrm{p}}\int\displaylimits_{a-c_\mathrm{0}t}^{b-c_\mathrm{0}t}  \tilde{\lambda}(\zeta ) \tilde{w}\left(\zeta +c_\mathrm{0}t\right) \mathrm{d}\zeta 	
\end{equation}
\end{subequations}
by substitution of \(\zeta = z - c_0 t\) and factorizing \(\lambda\left(\zeta\right)\) as the product of the constant term \(\lambda_{\mathrm{c}}\), which has the units of a line charge density, and a dimensionless longitudinal distribution function \(\tilde{\lambda}(\zeta)\). Despite the explicit dependence of the bunch form, the following derivation of the long bunch approximation only requires that the distribution \(\tilde{\lambda}(z)\) can be written as 
\begin{equation}\label{Eq:TaylorLineChargeDensity}
	\mathrm{T} \tilde{\lambda}(z; c_0 t) = \tilde{\lambda}(c_0 t) + \tilde{\lambda}^{\prime}(c_0 t)(z-c_0 t) + ... \ ,
\end{equation}
\(\forall z \in \mathbb{R}\) apart to a at most countable set of points, where \(\mathrm{T} \tilde{\lambda}(z; c_0 t)\) is the Taylor expansion of \(\tilde{\lambda}(z)\) at \(c_0 t\). The image charge can be written as an expansion as well,
\begin{subequations}
\begin{equation}
	Q_\mathrm{im}\left(t\right) = \frac{\lambda_{\mathrm{c}}}{2 \pi r_\mathrm{p}} \ \sum_{n=0}^{\infty} q_{n}(c_0 t) ,
\end{equation}
with 
\begin{equation}
	q_{n}(c_0 t) = \tilde{\lambda}^{(n)}(c_0 t) \int\displaylimits_{a-c_\mathrm{0}t}^{b-c_\mathrm{0}t}  \tilde{w}\left(\zeta+c_\mathrm{0}t\right) (\zeta-c_0 t)^{n} \mathrm{d}\zeta .              
\end{equation}
\end{subequations}
For a long bunch, or equivalently a short pickup, it is justified to consider only the leading term, with \(n = 0\), if the reminder term is negligible, which implies the function to be constant in the rolling interval of length \(\ell\). This is applicable if the following limit is true
\begin{equation}\label{Eq:ReqLongBunchApprox}
	\left|R_0\right| =	\left|\tilde{\lambda}(z) - \tilde{\lambda}(c_0 t)\right| \rightarrow 0 \ \ \ \forall z \in \left[c_0 t  - a, c_0 t +b\right] 
\end{equation}
and justified for any given function at some point for \(\left|a-b\right|\rightarrow 0\). Therefore, it is useful to introduce a scaling quantity \(\mu\) for the bunch-pickup system, which controls the approximation for any Lipschitz continuous function. For this purpose because the derivative is bounded by \(M_1 = \mathrm{sup}\left\{\left|\tilde{\lambda}^{\prime}(t)\right| : t \in \mathbb{R}\right\}\), that is the Lipschitz constant of the distribution \(\tilde{\lambda}(t)\), the following inequality gives an upper limit for the reminder \(R_0\)
\begin{equation}
	\tilde{\lambda}(z) - \tilde{\lambda}(c_0 t) \leq M_1 \left|z - c_0 t \right|  \leq M_1 \left|b - a \right| .
\end{equation}
Therefore, the scaling is defined as
\begin{equation}
	\mu = M_1 \left|b - a \right| = M_1 \ell
\end{equation}
and the requirement in Eq. \eqref{Eq:ReqLongBunchApprox} is fullfilled for
\begin{equation}\label{Eq:ScalingConditionLBApprox}
	\mu \ll 1 \ .
\end{equation}
Applied on a Gaussian bunch according to Eq. \eqref{eq:GaussianBunchProfile} it is
\begin{equation}
	\mu= \frac{1}{\sqrt{e}} \frac{l}{\sigma_{\mathrm{z}}}  \ll 1 \ .
\end{equation}
Under this assumption, the image charge is \(q_0\) and thus
\begin{equation}
	Q_\mathrm{im}^\mathrm{LBA}\left(t\right) = \frac{\lambda(t)}{2 \pi r_\mathrm{p}}\int_{a}^{b}  \tilde{w}\left(z\right) \mathrm{d}z \ ,  
\end{equation}
which by Eq. \eqref{Eq:PickupSurface-General} is
\begin{equation}\label{eq:LongBunchApprox}
	Q_\mathrm{im}^\mathrm{LBA}\left(t\right) = \frac{A_\mathrm{p}}{2 \pi r_\mathrm{p}} \lambda(t) .
\end{equation}
This approximation is widespread and for example found in \cite{Shafer-1990, Smith-1996}. Equation \eqref{eq:LongBunchApprox} is valid for any pickup which is short enough to neglect the change of line charge density along the pickup's length, which is ensured by requirement \eqref{Eq:ReqLongBunchApprox}, independent of pickup and bunch form. The validity of this situation, i.e. of the long-bunch approximation, is qualitatively controlled by condition \eqref{Eq:ScalingConditionLBApprox}. If this condition is satisfied, it is ensured that the requirement \eqref{Eq:ReqLongBunchApprox} is also met. Nonetheless Eq. \eqref{Eq:ScalingConditionLBApprox} is a stricter condition and may be untrue for some functions that still are sufficiently described by the LBA. These requirements are met in many facilities, thus the theory is well developed and calculations for the image current and output voltage discussed in various publications on beam instrumentation and in particular beam position monitors as \cite{Shafer-1990}.

\subsubsection{Short Bunch Approximation}

\begin{figure*}
	\includegraphics[width=1\columnwidth]{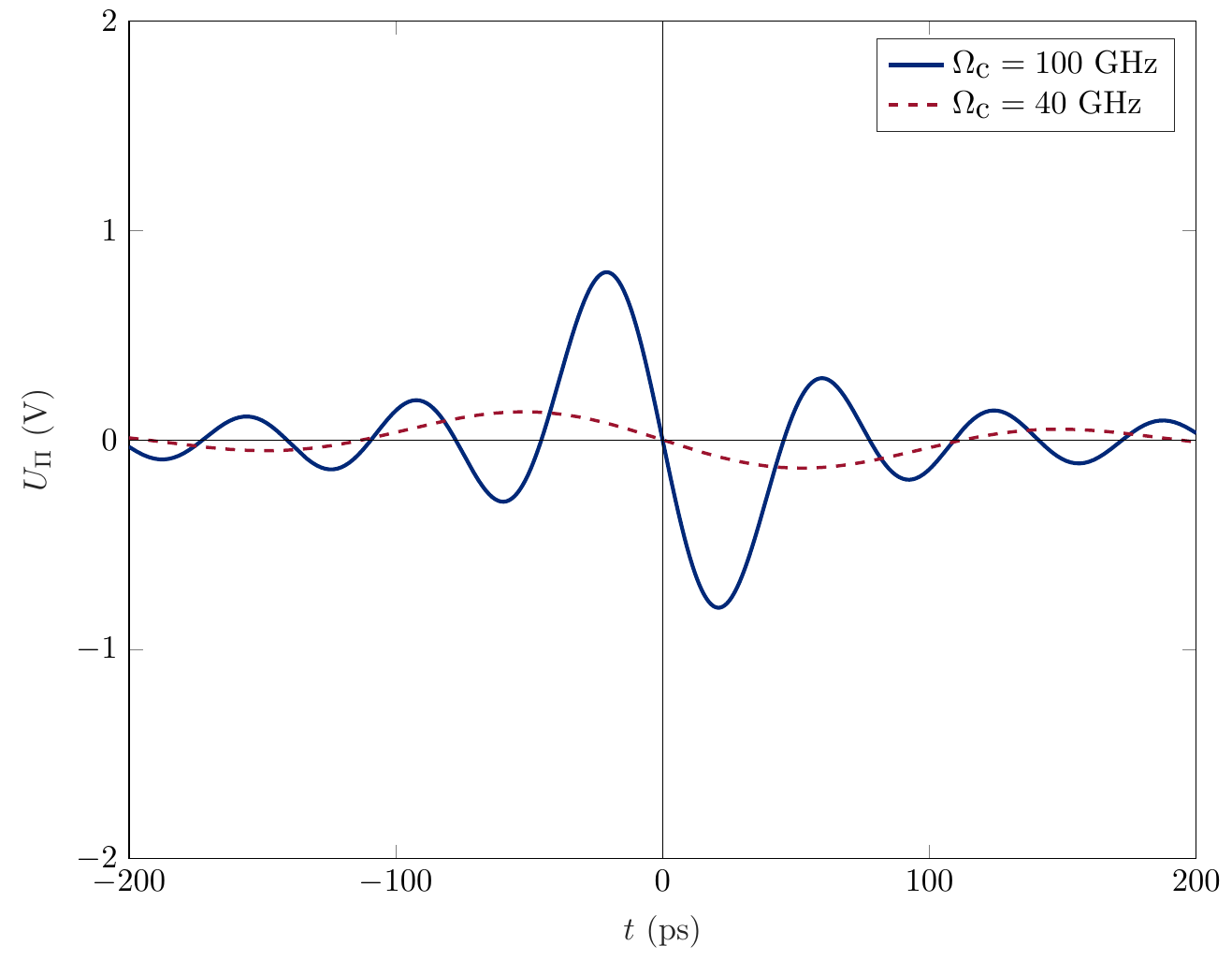}
	\hspace{0.05\columnwidth}
	\includegraphics[width=1\columnwidth]{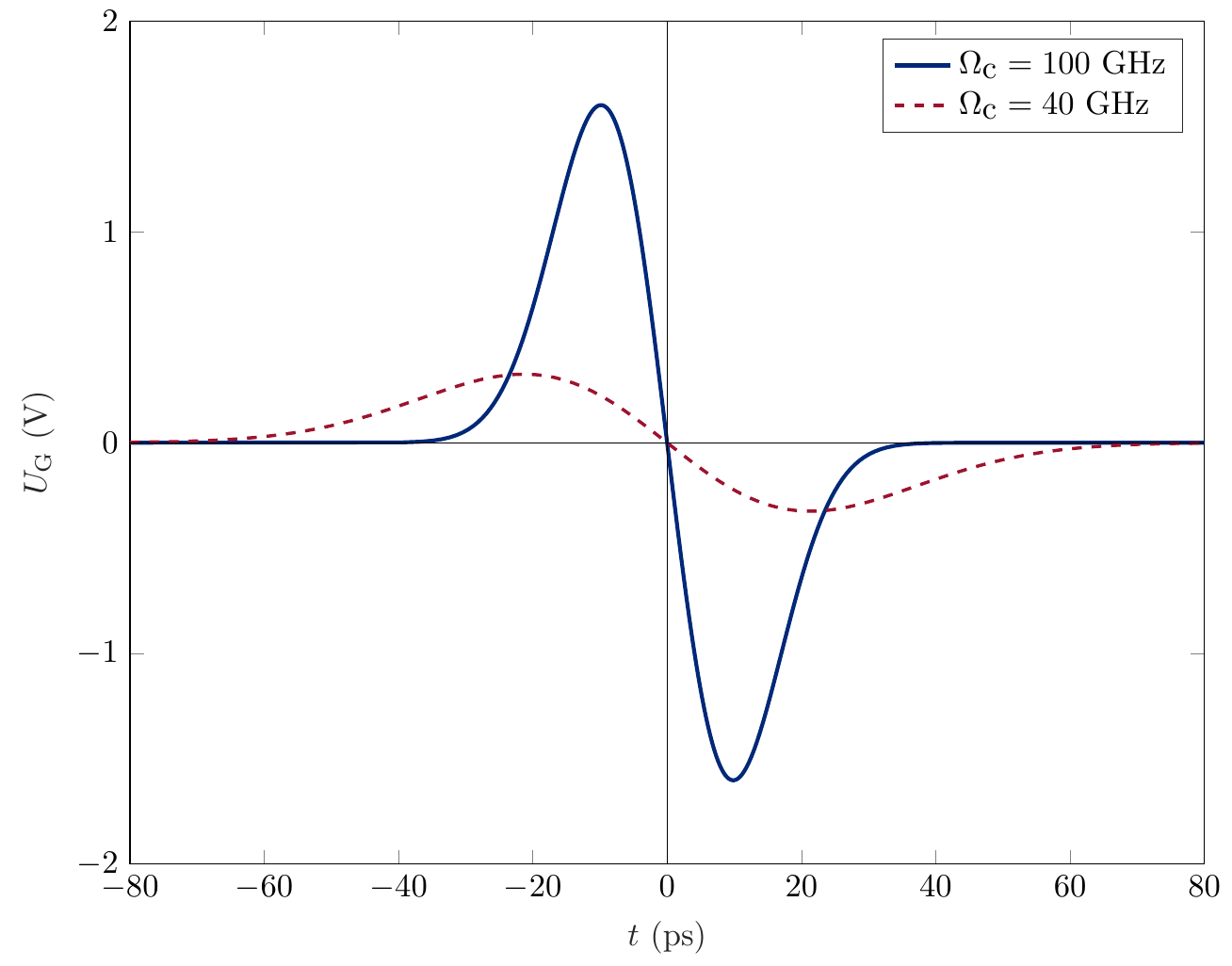}
	\caption{Voltage signal in the SBA for two idealized filters. A rectangular (left) and a Gaussian filter (right) have been utilized, each at \SI{100}{\giga\hertz} (blue) and \SI{40}{\giga\hertz} (red) cut-off frequency. Bunch and geometry parameters are in a reasonable range for XFEL applications, e.g. \(\sigma_{\mathrm{t}} = \SI{200}{\femto\second}\), \(Q_{\textrm{b}} = \SI{20}{\pico\coulomb}\), \(\ell = w_{0} = \SI{4.8}{\milli\meter}\), \(r_{\mathrm{p}} = \SI{20}{\milli\meter}\) and \(R = \SI{50}{\ohm}\). \label{fig:Voltage_InfinitesimalBunch1}}
\end{figure*}

In FEL applications as discussed in this work, the requirement of the long bunch approximation \(\mu \ll 1\) is not fulfilled. Instead the bunch length is typically below \SI{200}{\femto\second} down to single digit \si{\femto\second} and a standard pickup is in the range of some \si{\milli\meter}, corresponding to \(\geq \SI{3.3}{\pico\second}\). With the bunch more then one sometimes up to three orders of magnitude shorter it is worthwhile to investigate the limiting case of ultra short bunches, i.e. 
\begin{equation}\label{Eq:ScalingConditionSBApprox}
	\mu \gg 1 \ \Leftrightarrow \sigma_{\mathrm{z}} \ll \ell \ .
\end{equation}
Under this assumption in the short bunch approximation (SBA), because the normalized Gaussian is a Dirac sequence, Eq. \eqref{eq:GaussianBunchProfile} is described by a delta distribution with the total charge \(Q_\mathrm{b}\) concentrated at one point in space
\begin{equation}\label{eq:GaussianBunchProfileUltraShort}
	\lim\limits_{\sigma_{\mathrm{t}} \rightarrow 0} \lambda(z-c_\mathrm{0}t)
	= Q_{\mathrm{b}} \ \delta \left(z-c_0 t\right).
\end{equation}
Since the delta distribution is even, Eq. \eqref{eq:PUimageCharges} is treated as a convolution, leading straightforward to 
\begin{equation}\label{Eq:ShortBunch-Q-im}
	Q_\mathrm{im}^\mathrm{SBA}\left(t\right) = \frac{Q_{\mathrm{b}} w_0}{2 \pi r_\mathrm{p}}  \  \Pi_{2 t_0} \left( t \right) ,
\end{equation}
where 
\begin{equation}\label{Eq:Def-t0}
	t_0 = \frac{\ell}{2 c_0}
\end{equation}
is the time it takes for the bunch center to pass half the pickup. The image current found by differentiation of \({Q}_\mathrm{im}^\mathrm{SBA}(t)\) is given by the following distribution
\begin{equation}\label{eq:ultraShortBunchCurrent}
	I_\mathrm{im}^\mathrm{SBA}\left(t\right) = \frac{Q_{\mathrm{b}} w_0}{2 \pi r_\mathrm{p}} \left[\delta\left(t + t_0\right) - \delta\left(t - t_0\right)\right] .
\end{equation}
The output voltage is simply derived for any given transfer function \(H\left(\omega\right)\) with corresponding impulse response function \(h\left(t\right)\) by the convolution with Eq. \eqref{eq:ultraShortBunchCurrent}
\begin{equation}\label{Eq:STBasics-1}
	U_\mathrm{h}^\mathrm{SBA} \left(t\right) = \left(h \circledast I_\mathrm{im}\right) \left(t\right) .
\end{equation}
Therefore, the voltage signal in time domain is
\begin{equation}\label{Eq:VoltageSignal-ArbFilter-ShortBunch}
	U_\mathrm{h}^\mathrm{SBA} \left(t\right) = \frac{Q_{\mathrm{b}} w_0}{2 \pi r_\mathrm{p}} \left[h\left(t + t_0\right) - h\left(t - t_0\right)\right] .
\end{equation}
The preceding derivation of the short bunch limit is possible for all line charge density distributions in \textit{L}\textsuperscript{1}, thus with finite total bunch charge \(Q_\mathrm{b}\). The resulting voltage shape \(U_\mathrm{h}^\mathrm{s} \left(t\right)\) is certainly independent of the infinitesimal short bunch form, but defined by the subtraction of the scaled impulse response function \(h\left(t\right)\) shifted parallel in different directions.  

In the following two idealized transfer functions are used to illustrate the implications of the ultra-short bunch. The two applied low-pass filters are a rectangular filter defined by 
\begin{subequations}
	\begin{equation}
		H\left(\omega\right) = R \ \Pi_{2\Omega_{\mathrm{c}}} \left( \omega \right)
	\end{equation}
	and a Gaussian filter defined by
	\begin{equation}
		H\left(\omega\right) = R \exp \left(-\frac{\ln(2) \omega^2}{2  \Omega_{\mathrm{c}}^2}\right) .
	\end{equation}
\end{subequations}
The cut-off frequency is \(\Omega_{\mathrm{c}}\) and the filters are terminated with R, usually \SI{50}{\ohm}. The corresponding time domain response functions are
\begin{subequations}
\begin{equation}
	h_{\Pi}(t) = \frac{R \Omega_{\mathrm{c}}}{\pi} \sinci\left(\Omega_{\mathrm{c}} t\right)
\end{equation}
and
\begin{equation}
	h_{\mathrm{G}}(t) = \frac{R \widetilde{\Omega}_{\mathrm{c}}}{\sqrt{2 \pi}} \exp\left( - \frac{\widetilde{\Omega}_{\mathrm{c}}^2}{2} t^2\right) ,
\end{equation}
\end{subequations}
where \(\widetilde{\Omega}_{\mathrm{c}} = \Omega_{\mathrm{c}}/\sqrt{\ln (2)}\) is the cut-off frequency defined by a \(1/e\) decrease. The voltage signals \(U_\mathrm{\Pi}^\mathrm{s} \left(t\right)\) respectively \(U_\mathrm{G}^\mathrm{s} \left(t\right)\) are found by substitution of the response function in Eq. \eqref{Eq:VoltageSignal-ArbFilter-ShortBunch}. The resulting curves are shown in Fig. \ref{fig:Voltage_InfinitesimalBunch1} assuming dimensions in the range of existing bunch arrival time monitors. The characteristic curve, resembling the derivative of a Gaussian, is found at the center of both examples. An ideal rectangular filter induces an infinite ripple, which in some cases also reflects on the steep signal slope at the center, while the Gaussian is smooth. 

The signal slope at the zero-crossing
\begin{equation}
	S_{\mathrm{h,ZC}} = \dot{U}_{\mathrm{h}}\left(t_{\mathrm{ZC}}\right)
\end{equation}
is the figure of merit in BAM applications. The significant ZC is located at \(t=0\) for both example filters. The respective slopes are
\begin{subequations}
\begin{equation}
	S_{\mathrm{\Pi,ZC}}^\mathrm{SBA} = - \frac{R w_0}{\pi^2 r_{\mathrm{p}}} Q_{\mathrm{b}} \frac{\Omega_{\mathrm{c}}}{t_0} \left[
	\cos\left(\Omega_{\mathrm{c}} t_0\right) - \sinci\left(\Omega_{\mathrm{c}} t_0\right)	\right]
\end{equation}
and
\begin{equation}
	S_{\mathrm{G,ZC}}^\mathrm{SBA} = -\frac{R A_{\mathrm{p}}}{2 \pi r_{\mathrm{p}}} \frac{Q_{\mathrm{b}}}{\sqrt{2 \pi}} \frac{\widetilde{\Omega}_{\mathrm{c}}^3}{c_0}\left[\exp\left(-\frac{1}{2} \widetilde{\Omega}_{\mathrm{c}}^2 t_0^2\right)\right].
\end{equation}
\end{subequations}
For the rectangular filter the slope is depending on the ratio of cut-off frequency and pickup length, with a periodical behavior caused by the bracket term. This term has local extrema of approximately \(\pm 1\). For simplicity it is sufficient to assume exactly \(\pm 1\), found at \(\Omega_{\mathrm{c}} t_0 = n \pi\), with \(n \in \mathbb{N}\), although the real maximum is located somewhat before that. 

For the Gaussian filter the slope is also depending on the ratio of cut-off frequency and pickup length. For an infinite as well as infinitesimal pickup the slope is vanishing. The maximum slope is reached at \(\widetilde{\Omega}_{\mathrm{c}} = 2 c_0 / \ell\). For a lower cut-off frequency the signal width is increased, leading to distortion and partial annihilation of the signal. If the bandwidth is increased further, two distinctive peaks become visible leading to a lower gradient at zero-crossing. The optimum is
\begin{equation}\label{Eq:SGmax}
	S_{\mathrm{G,max}}^\mathrm{SBA} = - \frac{1}{\sqrt{e}} \frac{R A_{\mathrm{p}}}{2 \pi r_{\mathrm{p}}} \frac{Q_{\mathrm{b}}}{\sqrt{2 \pi}} \frac{1}{c_0} \widetilde{\Omega}_{\mathrm{c}}^3 .
\end{equation}
This equals the maximum slope received by the short pickup and high cut-off frequency approximation, apart from the factor \(1/\sqrt{e}\) and the replacement of the reciprocal bunch width by the cut-off frequency, which are the corresponding decisive widths.

\begin{figure*}
	\includegraphics[width=1\columnwidth]{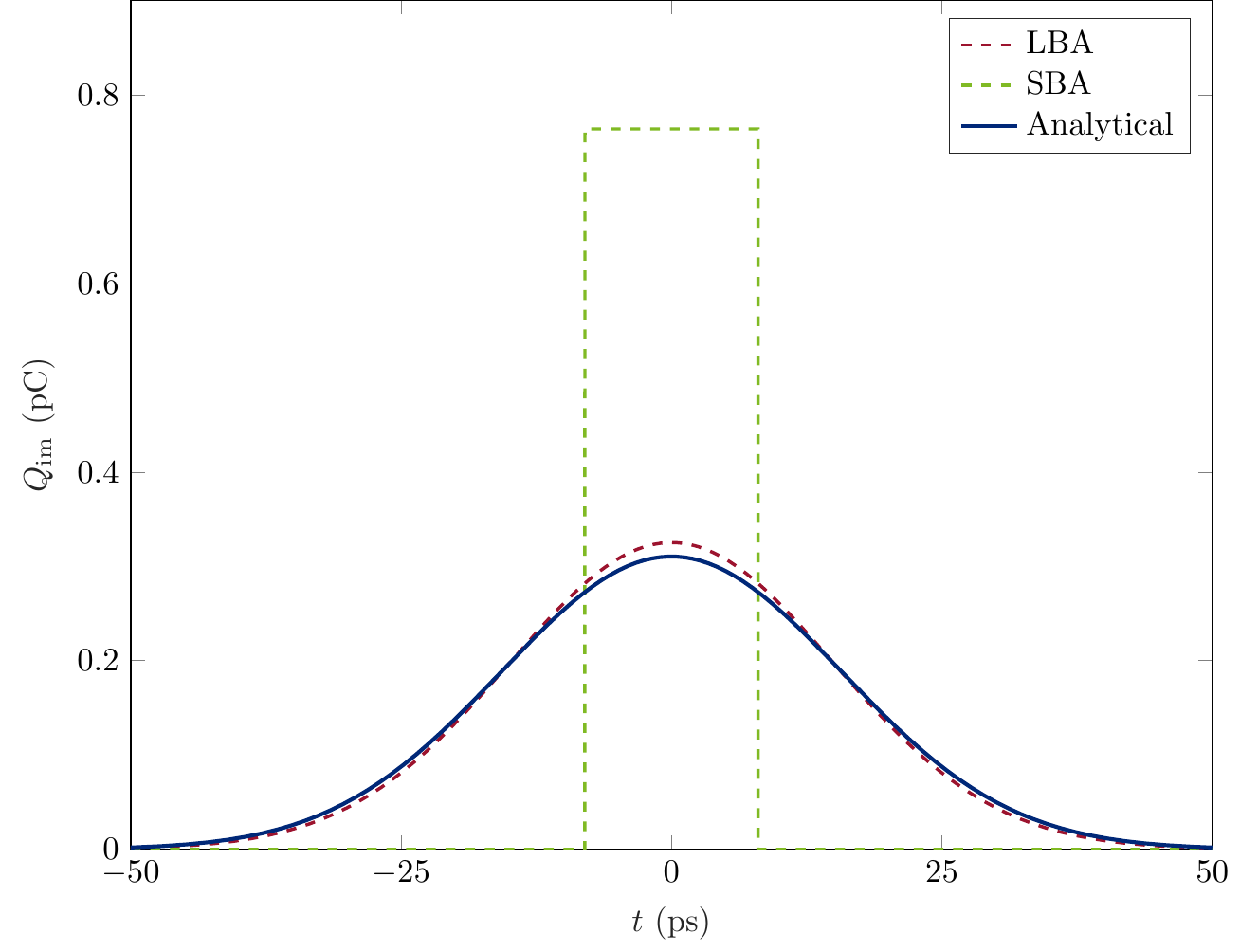}
	\hspace{0.07\columnwidth}
	\includegraphics[width=0.98\columnwidth]{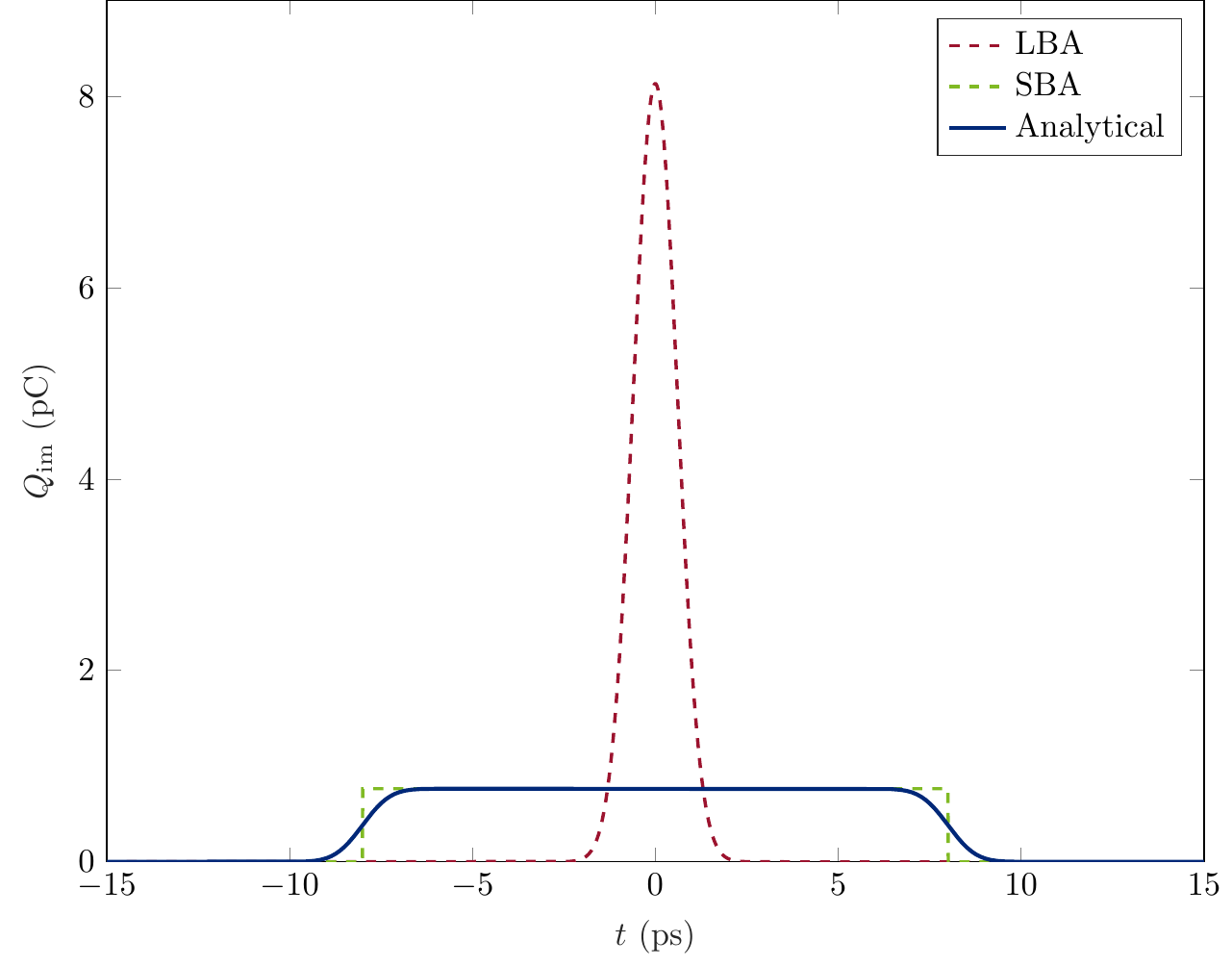}
	\caption{Image charge by a Gaussian bunch on a rectangular pickup surface according to the LBA in Eq. \eqref{eq:LongBunchApprox} (dashed red), the SBA in Eq. \eqref{Eq:ShortBunch-Q-im} (dashed green) and the analytical solution in Eq. \eqref{Eq:General-Q-im} (solid blue) for two different bunch lengths. One close to the LBA regime, with \(\mu \approx \num{0.65}\) (left) and one closer to the SBA regime, with \(\mu \approx \num{16.2}\) (right). Bunch and geometry parameters are in a reasonable range for applications of button pickups, e.g. \(\sigma_{\mathrm{t}} = \SI{0.6}{\pico\second}\) respectively \(\sigma_{\mathrm{t}} = \SI{15}{\pico\second}\), \(Q_{\textrm{b}} = \SI{20}{\pico\coulomb}\), \(\ell = w_{0} = \SI{4.8}{\milli\meter}\) and \(r_{\mathrm{p}} = \SI{20}{\milli\meter}\). In the EuXFEL even shorter bunches occur, further reducing the discrepancy between SBA and analytical solution at the edges. \label{fig:IMAGE_CURRENT}}
\end{figure*}

The real transfer function is neither rectangular nor Gaussian. In literature the pickup is usually modeled by an equivalent circuit of a resistor R parallel to the capacity C, with the image current \(I_\mathrm{im}\) serving as a current source \cite{Shafer-1990, Huang-2006}. The expected low-pass filter then gives
\begin{equation}\label{Eq:RC-Transfer-FD}
	H(\omega) = \frac{R}{1 + i \omega R C},
\end{equation}
with cut-off frequency \(\Omega_\mathrm{RC} = 1/\left(RC\right)\), as transfer function from current source \(I_\mathrm{im}\) to the output voltage \(U_\mathrm{RC}\). In time domain the impulse response function is
\begin{equation}\label{Eq:RC-Transfer-TD}
	h(t) = \frac{1}{C} \exp\left(- \frac{t}{R C}\right) \Theta\left(t\right) ,
\end{equation}
where \(\Theta\left(t\right)\) denotes the Heaviside step function.

In order to calculate the output voltage with the equivalent circuit's transfer function the convolution with the image current in short bunch approximation according to Eq. \eqref{Eq:VoltageSignal-ArbFilter-ShortBunch} gives
\begin{equation}
	\begin{aligned}
		U_\mathrm{RC}^\mathrm{SBA}(t) = \frac{1}{C} \frac{Q_{\mathrm{b}} w_0}{2 \pi r_\mathrm{p}} \exp\left(- \frac{t + t_0}{R C}\right) \\ 
		\times \left[\Theta\left(t + t_0\right) - \exp\left(+ \frac{2 t_0}{R C}\right) \Theta\left(t - t_0\right)\right] , 
	\end{aligned}
\end{equation}
which corresponds to the capacitor experiencing the complete electric field of the infinitesimal bunch at once, leading to a compensating current, which is equivalent to the discharging of a capacitor. The decay curve is observed until the coasting beam reaches the end of the pickup. Than the effect of the bunch's electric field on the pickup suddenly ends and the initial voltage jump is reversed. Afterwards the charges start flowing back, leading to a continuation of the exponential voltage decay. Therefore slope at zero-crossing, found at \(t = t_0\), is an infinite decrease.

\subsubsection{General Solution for the Image Current}
In case none of the preceding approximations is sufficient a general solution for Eq. \eqref{eq:PUimageCharges} is needed. It is found straightforward for Gaussian bunches and rectangular pickups. Applying the shapes defined in Eqs. \eqref{eq:GaussianBunchProfile} and \eqref{eq:RectangularPickup} without any approximation leads to
\begin{equation}
	Q_\mathrm{im}(t) = \frac{w_0 Q_\mathrm{b}}{2 \pi r_\mathrm{p}} \int_{-\ell/2}^{\ell/2} G\left(z-c_\mathrm{0}t, \sigma_\mathrm{z}\right)\mathrm{d}z .
\end{equation}
This integral is readily solved, giving
\begin{equation}\label{Eq:General-Q-im}
	Q_\mathrm{im}(t) = \frac{1}{2} \frac{w_0}{2 \pi r_\mathrm{p}} Q_\mathrm{b} \left[\erf\left( \frac{\frac{\ell}{2} - c_0 t}{\sqrt{2} \sigma_{\mathrm{z}}}\right)  +  \erf\left( \frac{\frac{\ell}{2} + c_0 t}{\sqrt{2} \sigma_{\mathrm{z}}}\right)\right]
\end{equation}
for the image charge induced on the pickup by the coasting beam. 

Figure \ref{fig:IMAGE_CURRENT} shows exemplary results of the analytical description in Eq. \eqref{Eq:General-Q-im} (blue) compared to the LBA, as in Eq. \eqref{eq:LongBunchApprox} (dashed red), and the SBA, as in Eq. \eqref{Eq:ShortBunch-Q-im} (dashed green), for two different values of the scaling factor \(\mu\). In case of an ultra short bunch, \(\mu \approx 16.2\) (right), when the coasting bunch is almost completely enclosed longitudinally by the pickup surface, a flat section is notable in the analytical result, which is in accordance with the SBA, while the LBA gives a sharp peak. In case of long bunches the curves found analytical and by LBA converge, but the SBA in Eq. \eqref{Eq:ShortBunch-Q-im} still gives a flat center with steep edges. Therefore both approximations are well suited for their cases, but differ significantly apart from that.

Differentiation of Eq. \eqref{Eq:General-Q-im} gives the general image current
\begin{equation}\label{Eq:GeneralImageCurrent}
 	I_\mathrm{im}(t) = \frac{w_0}{2 \pi r_\mathrm{p}} Q_b \left[G(t + t_0, \sigma_{\mathrm{t}})
 	 - G(t - t_0, \sigma_{\mathrm{t}})\right] ,
\end{equation}
with \(t_0\) defined in Eq. \eqref{Eq:Def-t0} and bunch width \(\sigma_{\mathrm{t}}\) expressed in units of time according to \eqref{Eq:Def-Sigt}. The first Gaussian represents the incoming charges introduced by the arriving bunch, the other is caused by the bunch leaving the effective area. This result can be understood as a generalization to the current found in the ultra short bunch approximation. If the delta distributions are replaced by Gaussians, the factor differs by the normalization \(1/\sqrt{2 \pi \sigma_{\mathrm{t}}^2}\). Consistently the general solution is readily converted into Eq. \eqref{eq:ultraShortBunchCurrent} by the limit of an ultra short bunch length.

\begin{figure*}
	\includegraphics[width=1.0\columnwidth]{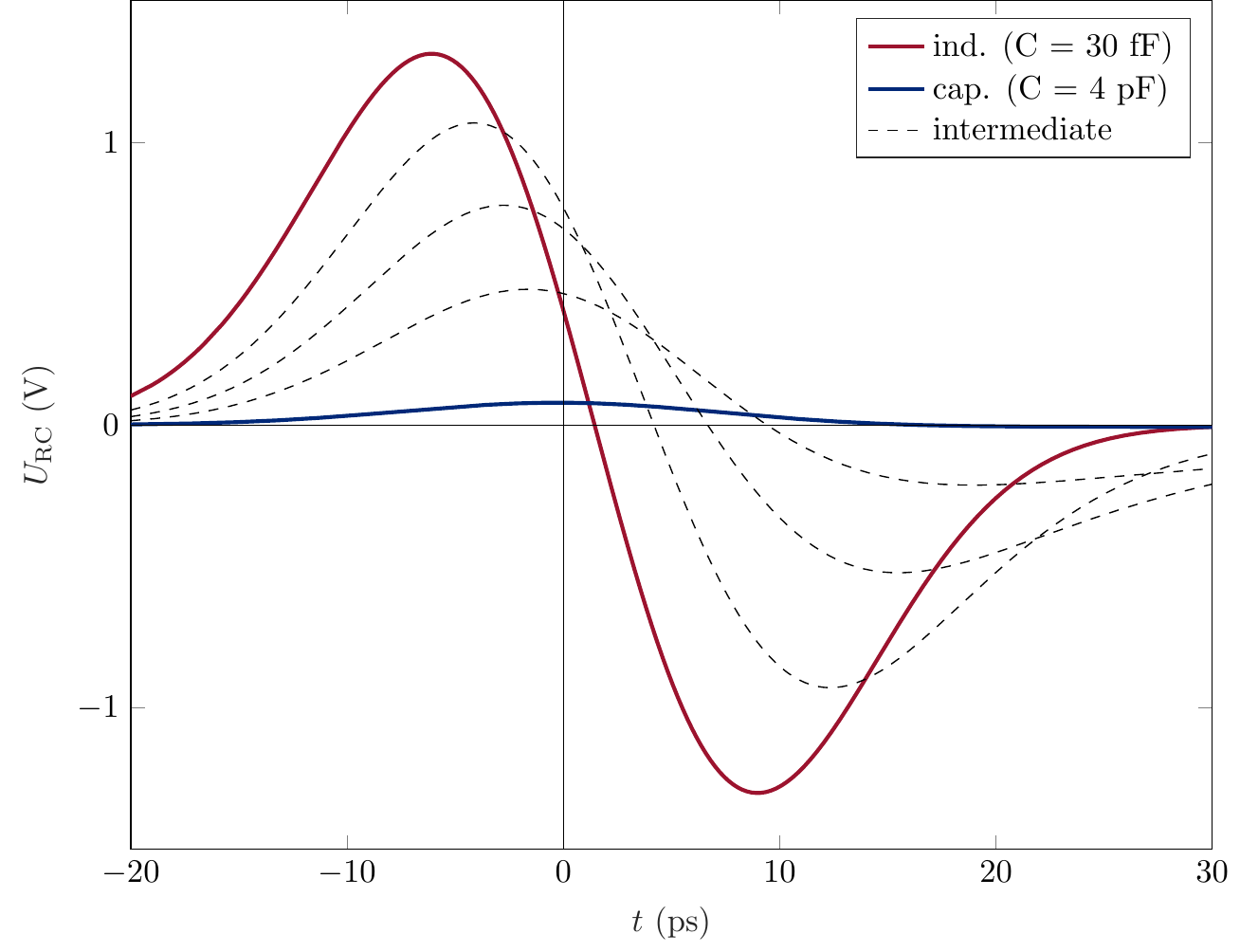}
	\hspace{0.05\columnwidth}
	\includegraphics[width=1.0\columnwidth]{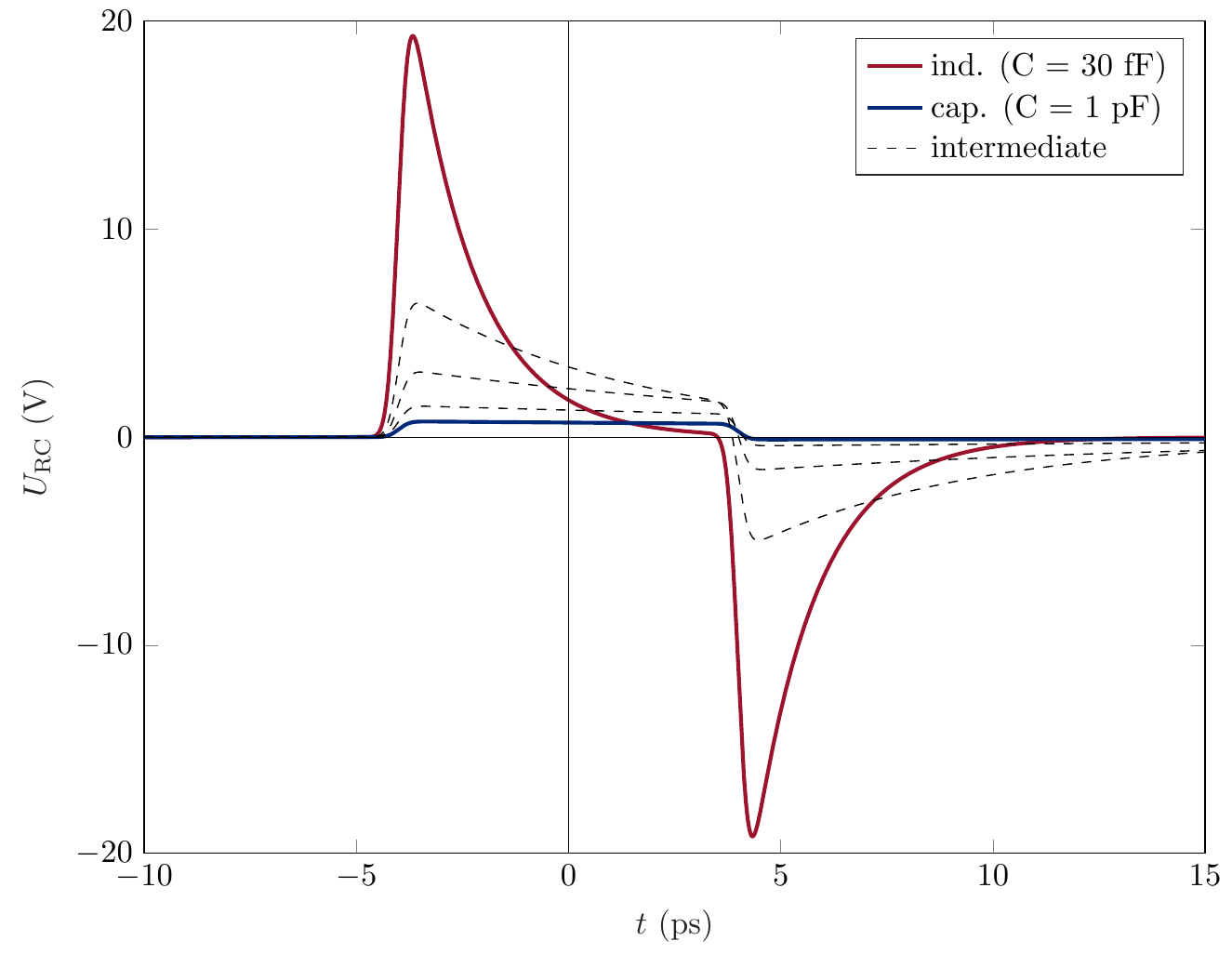}
	\caption{Voltage signal according to Eq. \eqref{Eq:VoltageSignal-Gerneral} with dimensions in the range of the LBA with \(\mu \approx 0.7\) (left) and SBA with \(\mu \approx 24\) (right). The capacity is varied from almost the inductive limit (red) up to the capacitive case (blue) with three intermediate steps (dashed). Bunch and geometry parameters are in a reasonable range for XFEL applications, e.g. \(\sigma_{\mathrm{t}} = \SI{7}{\pico\second}\) (left) respectively \(\sigma_{\mathrm{t}} = \SI{0.2}{\pico\second}\) (right), \(Q_{\textrm{b}} = \SI{20}{\pico\coulomb}\), \(\ell = \SI{2.4}{\milli\meter}\), \(w_{0} = \SI{4.8}{\milli\meter}\) and \(r_{\mathrm{p}} = \SI{20}{\milli\meter}\). In other applications, when the capacitive case is required, the limiting case can be achieved by increasing the terminating resistor R, to counter the \(1/C\) dependence in Eq. \eqref{Eq:VoltageSignal-Gerneral}, see Fig. 3 (a) in \cite{Huang-2006}.\label{fig:VoltageExamples}}
\end{figure*}

The general voltage signal measured at resistor R is found by the known transfer function based on the pickups equivalent circuit, see Eq. \eqref{Eq:RC-Transfer-FD}. It is determined by the convolution of the general image current, Eq. \eqref{Eq:GeneralImageCurrent}, and the impulse response function, Eq. \eqref{Eq:RC-Transfer-TD}. The result, also mentioned in \cite{Huang-2006}, is 
\begin{widetext}
\begin{equation}\label{Eq:VoltageSignal-Gerneral}
	\begin{aligned}
	U_{\mathrm{RC}}(t) = \frac{1}{2} \frac{1}{C} \frac{w_0}{2 \pi r_\mathrm{p} } Q_{\mathrm{b}}
	\exp\left(\frac{1}{2} \left(\frac{\sigma_{\mathrm{t}}}{R C}\right)^2 \right) 
	\left\{ 
	\exp\left(- \frac{t + t_0}{R C}	\right) \left[1 + \erf\left(\frac{t+t_0}{\sqrt{2}\sigma_{\mathrm{t}}}-\frac{\sigma_{\mathrm{t}}}{\sqrt{2}RC}\right)  \right] \right. -\\
	\left. \exp\left(- \frac{t - t_0}{R C}	\right) \left[1 + \erf\left(\frac{t-t_0}{\sqrt{2}\sigma_{\mathrm{t}}}-\frac{\sigma_{\mathrm{t}}}{\sqrt{2}RC}\right)  \right]
	\right\} .
	\end{aligned}
\end{equation}
\end{widetext}
The voltage curve shows a very different behavior for different parameter choice, which is not immediately accessible in Eq. \eqref{Eq:VoltageSignal-Gerneral}. The two introduced limiting cases for \(\mu\ll 1\) and \(\mu \gg 1\) are in accordance with the general solution. For an intermediate value of \(\mu\), it is possible to find an approximation of Eq. \eqref{Eq:VoltageSignal-Gerneral} by assuming extreme values for the cut-off frequency or in particular the capacity, as presented in the next section. The limiting cases and the transitional behavior are exemplified in Fig. \ref{fig:VoltageExamples}. On the left side bunch and pickup length approach the LBA, \(\mu \approx 0.7\). In contrast the right image contains signals for a relatively short bunch with \(\mu \approx 24\), looking alike the SBA. In both plots the capacitance is varied starting at \SI{30}{\femto\farad}, close to the so called inductive case (red) to \SI{1}{\pico\farad} respectively \SI{4}{\pico\farad}, which is around the capacitive case (blue). The corresponding cut-off frequencies are \(\Omega_{\mathrm{RC}} = \SI{0.67}{\tera\hertz}\) and \SI{5}{\giga\hertz} respectively \SI{20}{\giga\hertz}. The transition between both cases is indicated with stripped lines.

The BAM resolution depends on the signal slope at the zero-crossing (ZC). The time dependent signal slope found by derivation of Eq. \eqref{Eq:VoltageSignal-Gerneral}
\begin{equation}\label{Eq:SigSlope-General}
	\dot{U}_{\mathrm{RC}}(t) = \Omega_\mathrm{RC} \left[ R I_\mathrm{im}(t) - U_{\mathrm{RC}}(t) \right].
\end{equation}
At the time of zero-crossing \(t_{\mathrm{ZC}}\) the voltage \(U_{\mathrm{RC}}(t_{\mathrm{ZC}})\) is zero by definition. Therefore, the signal slope at the zero-crossing is only defined by the image current
\begin{equation}\label{Eq:SigSlope-General-I}
	S_{\mathrm{RC,ZC}} = \dot{U}_{\mathrm{RC}}(t_{\mathrm{ZC}}) = \frac{1}{C} I_\mathrm{im}(t_{\mathrm{ZC}}).
\end{equation}
Because \(I_\mathrm{im}(t=0)\) is zero, a zero-crossing at \(t_{\mathrm{ZC}} = 0\) would entail a gradient of zero, but the zero-crossing is located at \(t_{\mathrm{ZC}} \neq 0\) for any \(C > 0\).

\subsubsection{Voltage by Frequency Regions}

In many theoretical approaches the voltage signal is analyzed in two special cases defined by frequency regions. In the so called capacitive limit, the capacity C is large, respectively the cut-off frequency approaches zero \(\Omega_\mathrm{RC} \rightarrow 0\), hence all exponential functions approach one and the first term of each error function is predominant. This leads to a great simplification of the output voltage, which is
\begin{equation}
	U_{\mathrm{RC}}^\mathrm{cap}(t) = \frac{1}{C} Q_{\mathrm{im}}(t) .
\end{equation}
Therefore, the signal in the limit of a infinitesimal cut-off frequency resembles a charge source \cite{Huang-2006}. The limit for an infinitesimal capacity, the inductive case \cite{Huang-2006} with \(\Omega_\mathrm{RC} \rightarrow \infty\), is straightforward for the frequency domain transfer function Eq. \eqref{Eq:RC-Transfer-FD}, which becomes the constant R, and thus \(h(t) = R \delta(t)\) in time domain. The voltage is then
\begin{equation}\label{Eq:InductiveLimitVoltage}
	U_{\mathrm{RC}}^\mathrm{ind}(t) = R I_{\mathrm{im}}(t) , 
\end{equation}
resembling a current source \cite{Huang-2006}. In this case, the zero-crossing is exactly at \(t=0\) and for a finite pickup length the signal slope is nonzero in a seeming contradiction to Eq.  \eqref{Eq:SigSlope-General-I}, but the denominator C is zero as well.

\subsubsection{Maximum Voltage and Signal Slope}

For deployment in a BAM, maximum voltage and signal slope are the key features of the pickup. Moreover a throughout understanding of the decisive parameters is crucial. The general formulation in Eq. \eqref{Eq:VoltageSignal-Gerneral} confirms the proportionality to the bunch charge and pickup width as well as reciprocal to the distance between bunch and pickup 
\begin{equation}
	U_{\mathrm{RC,max}} \propto Q_{\mathrm{b}}\frac{w_0}{r_{\mathrm{p}}}
\end{equation}
and
\begin{equation}\label{Eq:Slope-Dependencies}
		S_{\mathrm{RC,max}} \propto Q_{\mathrm{b}} \frac{w_0}{r_{\mathrm{p}}} .
\end{equation} 
The proportionality to the bunch charge was experimentally proven for the signal slope of the cone shaped pickups at FLASH \cite{Angelovski-2015}. The 
inverse proportionality to \(r_\mathrm{p}\) has to be analyzed in simulations, since these measurements are expensive and limited by the facilities design parameters. 

Another dependency observed in simulations, is on the ratio of pickup and bunch lengths. The maximum voltage approaches its maximum asymptotically and deviates only slightly from the value for a pickup longer than a few bunch lengths \cite{Huang-2006, Angelovski-2013}. This also effects the signal slope, as the maximum voltage barely changes in this region while both extrema drift apart, leading to a decreasing slope \cite{Angelovski-2013}. In Eq. \eqref{Eq:VoltageSignal-Gerneral} it is readily shown, that the voltage vanishes for \(t_0 = 0\), which is identical to \(\ell = 0\). To describe the behavior for any other ratio, the approximations must be utilized. In case of the pure charge source the maximum voltage found at \(t_{\mathrm{max}} = 0\) is
\begin{equation}\label{Eq:Vmax-cap}
	\max\left(U_{\mathrm{RC}}^{\mathrm{cap}}\right) =\frac{1}{C} \frac{w_0  Q_\mathrm{b}}{2 \pi r_\mathrm{p}} \erf\left( \frac{\ell}{\sqrt{8} \sigma_{\mathrm{z}}}\right)
\end{equation}
\cite{Huang-2006}. For a pure current source \(t_{\mathrm{max}}\) must adhere to
\begin{equation}
	t_0 = t_{\mathrm{max}} \tanh \left(\frac{t_0}{\sigma_{\mathrm{t}}} \frac{t_{\mathrm{max}}}{\sigma_{\mathrm{t}}}  \right), 
\end{equation}
which is found by setting \(\dot{U}_{\mathrm{RC}}^{\mathrm{ind}} = 0\). Two useful approximations give the positions of extrema for a large or negligible argument of the hyperbolic tangent, which are \(t_{\mathrm{max}} \approx \pm \sigma_{\mathrm{t}}\) for the LBA with \(t_0 \ll \sigma_{\mathrm{t}}\) respectively \(t_{\mathrm{max}} \approx \pm t_0\) for the SBA with \(t_0 \gg \sigma_{\mathrm{t}}\). The corresponding maximum voltages are for a short pickup
\begin{equation}\label{Eq:Vmax-ind-shortPU}
	\max\left(U_{\mathrm{RC,t_0 \ll \sigma_{\mathrm{t}}}}^{\mathrm{LBA,ind}}\right) \approx 
	\frac{R}{\sqrt{e}}
	\frac{A_{\mathrm{p}}}{2 \pi r_\mathrm{p}} 
	\frac{Q_\mathrm{b}}{\sqrt{2 \pi} c_0 \sigma_{\mathrm{t}}^2} = \mu \hat{U}_{0}^{\mathrm{ind}}, 
\end{equation}
with 
\begin{equation}
	\hat{U}_{0}^{\mathrm{ind}} = R \frac{w_0}{2 \pi r_\mathrm{p}}  \frac{Q_\mathrm{b}}{\sqrt{2 \pi}\sigma_{\mathrm{t}}}, 
\end{equation}
and for a long pickup
\begin{equation}\label{Eq:Vmax-ind-longPU}
	\max\left(U_{\mathrm{RC,t_0 \gg \sigma_{\mathrm{t}}}}^{\mathrm{SBA,ind}}\right) \approx 
	\hat{U}_{0}^{\mathrm{ind}} \left[1 - \exp\left(-\frac{l^2}{2 \sigma_{\mathrm{z}}^2}\right)\right].  
\end{equation}
The first term is the maximum voltage for \(\ell \rightarrow \infty\), which is justified for \(\ell > 3 \sigma_{\mathrm{z}}\) with the Gaussian bunch. In the cases described by Eqs. \eqref{Eq:Vmax-cap} and \eqref{Eq:Vmax-ind-longPU} the maximum voltage saturates for a pickup length of a few \(\sigma_{\mathrm{z}}\), whereas in the realms of LBA, with an infinitesimal pickup, the maximum voltage goes linear with its length.

\begin{figure}
	\includegraphics[width=\columnwidth]{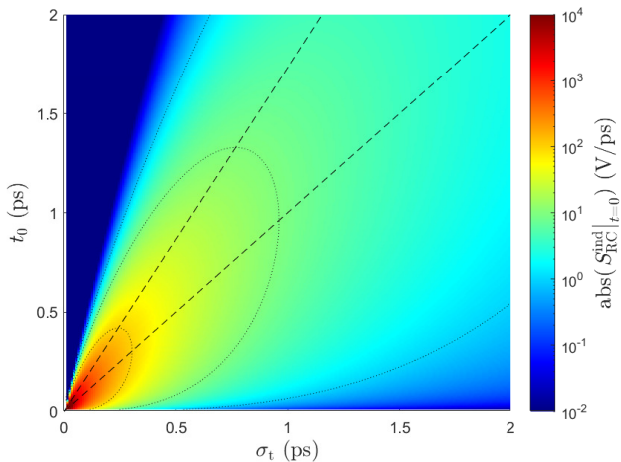}
	\caption{Absolute signal slope at zero-crossing according to Eq. \eqref{Eq:ZCSlopeIndRC} as a function of \(t_0\) and \(\sigma_{\mathrm{t}}\) in the interval from \SIrange[range-units = single]{0}{2}{\pico\second}. The dotted contours mark \num{1}, \num{10} and \SI{100}{\volt\per\pico\second} and dashed lines indicate \(t_0 = \sigma_{\mathrm{t}}\) and \(t_0 = \sqrt{3} \sigma_{\mathrm{t}}\).\label{fig:OptimalLengthRatio}}
\end{figure}

Regarding the BAM resolution a high signal slope is desired. The main proportionality is already given by Eq. \eqref{Eq:Slope-Dependencies}, but in the inductive limit a general solution is possible. The slope at the zero-crossing is
\begin{equation}\label{Eq:ZCSlopeIndRC}
	\left. S^{\mathrm{ind}}_{\mathrm{RC}} \right|_{t=0} = - R Q_{\mathrm{b}} \frac{w_0}{\pi r_\mathrm{p}} \frac{t_0}{\sigma_{\mathrm{t}}^2}  G\left(t_0,\sigma_{\mathrm{t}}\right) .
\end{equation}
This result is visualized in Fig. \ref{fig:OptimalLengthRatio}. For a fixed bunch length, the optimum is reached at \(t_0 = \sigma_{\mathrm{t}}\), whereas for a fixed pickup length the best result is found at \(t_0 = \sqrt{3} \sigma_{\mathrm{t}}\). At \(t_0 = \sigma_{\mathrm{t}} = 0\) is a pole, therefore reduction of both dimensions is favorable in the case of a pure charge source, but in real applications the bunch length is in a fixed range specific for the facility and foreseen experiments.

\subsubsection{Bandwidth Limitation}

The measured signal slope is limited by the bandwidth according to the uncertainty principle. The product of a signal's rise time \(\tau\) as response to a step function and the frequency bandwidth \(\Delta f\) satisfies\footnote{Denote that the right-hand side of the related uncertainty inequalities by K{\"u}pfm{\"u}ller \cite{Kupfmuller-1932} and Gabor \cite{Gabor-1946} depend on the definitions of bandwidth and duration.}
\begin{equation}\label{eq:KupfmullerUP}
	\tau \Delta f = 1 .
\end{equation}
This rise time serves as an upper limit and can be expressed by the maximum slope \(S_\mathrm{max}\) and the peak-to-peak voltage \(U_\mathrm{PP}\) with \(\tau = U_\mathrm{PP} / S_\mathrm{max}\) giving
\begin{equation}\label{eq:KupfmullerUP_Slope}
	\begin{aligned}
		S_\mathrm{max} 	& \leq U_\mathrm{pp} \ \Delta f .
	\end{aligned}
\end{equation}

This limitation affects the resolution of a transmitted signal. It determines the highest possible slope allowed in the response function, but if the slope of an incoming signal is well below this limit, a rise in bandwidth might even have a negative effect on the outputs signal slope, as already shown for the idealized Gaussian filter in the SBA.

\subsubsection{Circular Pickup}

In all prior considerations the pickup was assumed rectangular, which is not in accordance with many real-world applications. From Fig. \ref{fig:VoltageExamples} and the formulation of the SBA it is reasonable to assume that the rectangular model is sufficient for long bunches, but its shape is significantly reflected in the voltage signal for shorter bunches. A general analytical solution for the circular surface of radius \(r_{\mathrm{B}}\), defined by
\begin{equation}\label{eq:RoundPickup}
	w(z) = 
	\begin{cases}
		2 \sqrt{r_{\mathrm{B}}^2 - z^2} 	&|z| \leq r_{\mathrm{B}}\\
		0 									&|z| > r_{\mathrm{B}}
	\end{cases}	
\end{equation}
\cite{Smith-1996}, is not available and numerical methods have to be utilized. Nonetheless, some special cases are accessible by analysis. For the SBA, according to Eq. \eqref{eq:GaussianBunchProfileUltraShort}, the image charge is proportional to the pickup shape and thus the image current is
\begin{equation}
	I_{\mathrm{im}}^{\mathrm{SBA}}(t) \approx
	\begin{cases}
		- \frac{c_0^2 Q_{\mathrm{b}}}{\pi r_{\mathrm{p}}} \frac{t}{\sqrt{r_{\mathrm{B}}^2 - \left(c_0 t\right)^2}} 	&|t| < \frac{r_{\mathrm{B}}}{c_0}\\
		0 									&|t| > \frac{r_{\mathrm{B}}}{c_0}
	\end{cases}	
\end{equation}
with two poles at \(t_{\pm\infty} = \pm r_{\mathrm{B}}/c_0\), which would be softened by a broader bunch. In the inductive limit, interesting for synchronization, the voltage is given by Eq. \eqref{Eq:InductiveLimitVoltage}, with the zero-crossing exactly at \(t = 0\) and a signal slope of
\begin{equation}
	\left. S_{\mathrm{RC}}^{\mathrm{SBA,ind}} \right|_{t=0} = -R \frac{c_0^2 Q_{\mathrm{b}}}{\pi r_{\mathrm{p}}} \frac{1}{r_{\mathrm{B}}} .
\end{equation}
The slope is inversely proportional to the radius of the button pickup, hence a smaller button is favorable, though this approximation is only valid for a button much larger than the bunch.

\subsubsection{Summary}

An analytical solution for the voltage signal is readily found, which is valid for any bunch or pickup length and for any value of the lumped elements in the equivalent circuit representing the physical pickup. For different limiting cases, as the SBA, LBA, capacitive as well as inductive limit, appropriate expressions exist, which are easier to work with.

Nonetheless, some evaluations were carried out for the general solution as well. First of all the highly relativistic bunch is assumed Gaussian and centered on the longitudinal axis. Usually the pickup was treated as a rectangular surface with the same curvature as the beam pipe, to keep the radial distance to the beam constant. Fringing fields caused by the gap between pickup surface and the beam pipe are neglected, but might be considered sufficiently by a constant factor depending on the geometry. The equivalent circuit was idealized as RC elements in parallel.

The most critical approximation is the rectangular pickup surface. The solution is still useful in the range of medium to long bunches, but the form becomes a significant factor for \(\mu \gg 1\). There are numerical methods for arbitrary pickup surfaces.

\subsection{Numerical Solution}

\begin{figure*}
	\includegraphics[width=\columnwidth]{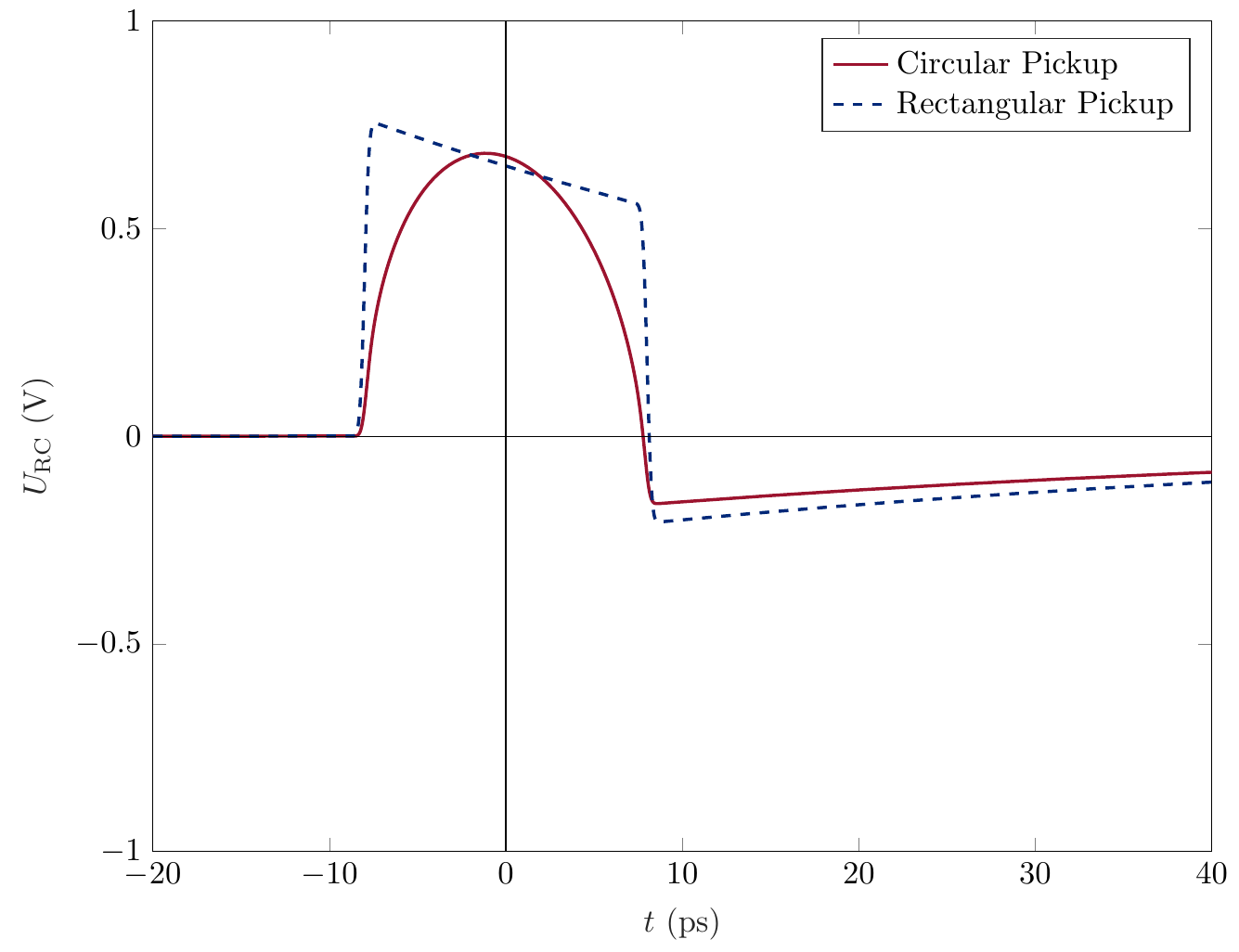}
	\hspace{0.05\columnwidth}
	\includegraphics[width=\columnwidth]{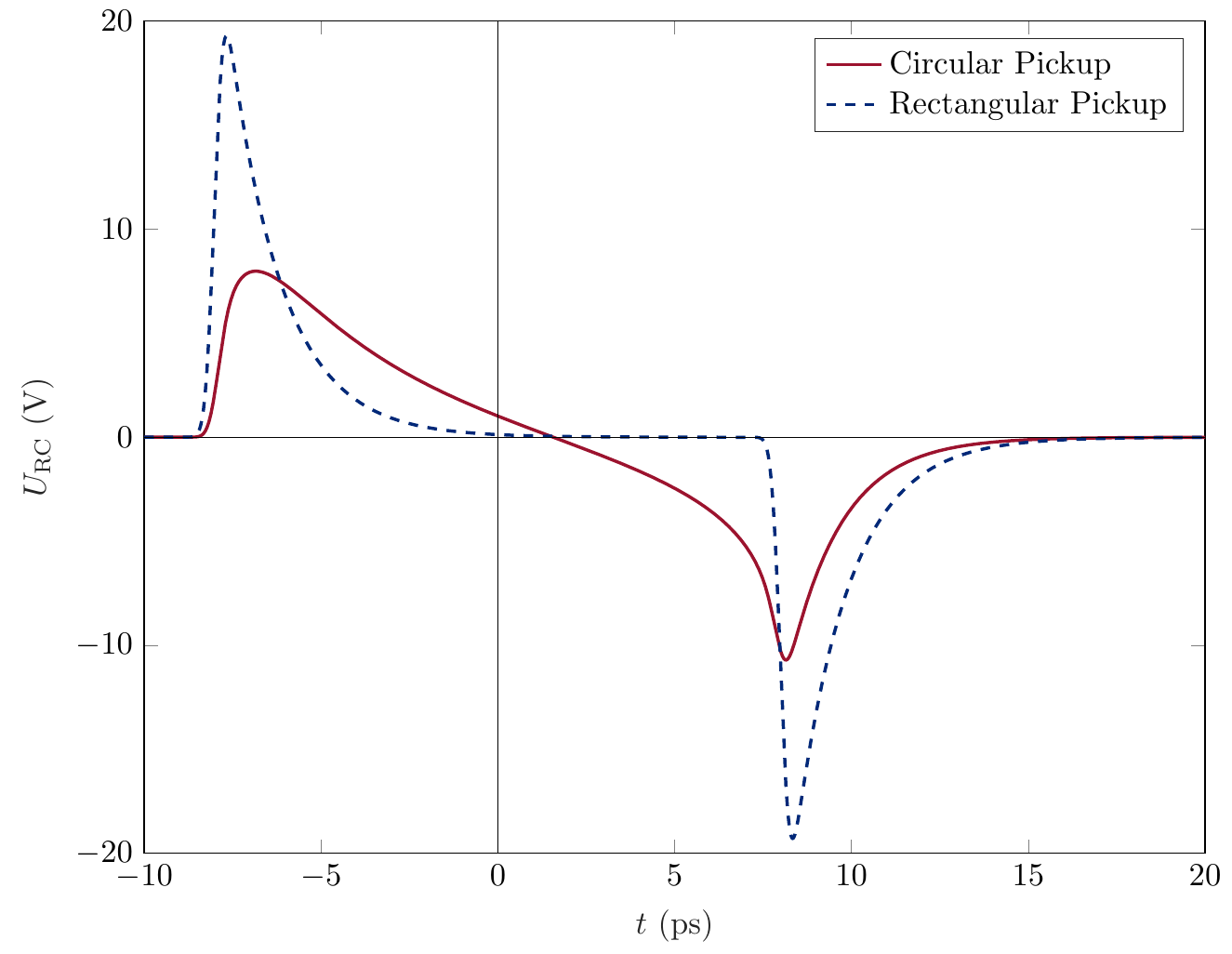}
	\caption{Voltage signal of a rectangular (dashed blue) and a round (solid red) pickup surface calculated with Matlab\textsuperscript{\textregistered} according to Eqs. \eqref{Eq:NumQim-FD} to \eqref{Eq:NumVo-TD}. Analog to the analytic approach, a Gaussian bunch and an RC-filter were used. The dimensions are in range of the SBA with \(\mu \approx 49\) and at the capacitive limit (left) respectively inductive limit (right). Bunch and geometry parameters, apart from the pickup length, are the same as in Fig. \ref{fig:VoltageExamples} (right), e.g. \(R = \SI{50}{\ohm}\), \(C = \SI{1}{\pico\farad}\) (left) respectively \(C = \SI{30}{\femto\farad}\) (right), \(\sigma_{\mathrm{t}} = \SI{0.2}{\pico\second}\), \(Q_{\textrm{b}} = \SI{20}{\pico\coulomb}\), \(\ell = w_{0} = r_{\mathrm{B}} = \SI{4.8}{\milli\meter}\) and \(r_{\mathrm{p}} = \SI{20}{\milli\meter}\).\label{fig:NumericalSolution}}
\end{figure*}

A numerical solution of Eq. \eqref{eq:PUimageCharges} is accessible for any bunch or pickup shape, to study the signal in cases where the prior assumptions are not valid. While the Gaussian bunch in Eq. \eqref{eq:GaussianBunchProfile} usually is a good approximation, the button pickup has a circular surface with a width given by Eq. \eqref{eq:RoundPickup}. In Fig. \ref{fig:VoltageExamples} the pickup form is apparent in the signal shape in case of short bunches. An analytical solution was only determined in the SBA with an inductive pickup. It is therefore crucial to find a general description considering the pickup form, which is possible by numerical methods. Acknowledging that for any even charge distribution, Eq. \eqref{eq:PUimageCharges} corresponds to a convolution, the image charge is readily calculated in frequency domain by the convolution theorem. It is
\begin{equation}\label{Eq:NumQim-FD}
	Q_\mathrm{im}\left(\omega\right) = \mathrm{FFT}\left[\frac{\lambda(c_0 t)}{2 \pi r_{\mathrm{p}}}\right] \mathrm{FFT}\left[w(c_0 t)\right], 
\end{equation}
where \(\mathrm{FFT}\) is the fast fourier transform. In case of an uneven charge density distribution it is necessary to exchange \(\mathrm{FFT}\left[\lambda(c_0 t)\right]\) by the complex conjugate \(\overline{\mathrm{FFT}\left[\lambda(c_0 t)\right]}\). By inverse FFT the time domain image charge is
\begin{equation}\label{Eq:NumQim-TD}
	Q_\mathrm{im}\left(t\right) = \mathrm{FFT}^{-1}\left[Q_\mathrm{im}\left(\omega\right) \right] 
\end{equation}
\cite{Smith-1996}. The image current is 
\begin{equation}\label{Eq:NumIim-TD}
	I_\mathrm{im}\left(t\right) = \mathrm{FFT}^{-1}\left[i \omega Q_\mathrm{im}\left(\omega\right) \right]  
\end{equation}
and the voltage at the output of a system with transfer function \(H(\omega)\) 
\begin{equation}\label{Eq:NumVo-TD}
	U_\mathrm{h}\left(t\right) = \mathrm{FFT}^{-1}\left[H(\omega)  I_\mathrm{im}\left(\omega\right) \right] .
\end{equation}

This method works well for finite inputs. Residual values, e.g. of the response function, at the end of the time interval lead to unphysical effects. Therefore, the time interval and sample frequency must be chosen with care.

The voltage signal calculated with a parameter set in the range of the operational parameters at the EuXFEL is pictured in Fig. \ref{fig:NumericalSolution}. The dashed blue line represents the numerical result for a rectangular pickup, which is in good agreement with the analytical version in Fig. \ref{fig:VoltageExamples} (right). The red line shows the signal by a circular pickup.

\section{Ultra-low charge mode}
For new experiments in the EuXFEL, with an ultra-low charge mode (\(\leq\)\SI{1}{\pico\coulomb}), the 2\textsuperscript{nd} generation pickups are incapable of providing the required signal for \si{\femto\second} resolution. Therefore, the BAM will be upgraded with a novel pickup structure and new EOM.

\subsection{Planned signal improvement}
New layouts are restrained in each facility by design regulations and previous design choices. For the next BAM upgrade in the EuXFEL a smaller pipe diameter is now permitted. The possible reduction from \SI{40.5}{\milli\meter} \cite{Angelovski-2012} to \SI{10}{\milli\meter} gives a potentially fourfold signal increase, by relation \eqref{Eq:Slope-Dependencies}. When additionally the bandwidth is raised, according to inequality \eqref{eq:KupfmullerUP_Slope}, one can expect a total improvement by one order of magnitude. Moreover, it is possible to combine multiple signals and to shorten the lossy RF path for further improvement.

The downside of these changes is a shorter dynamical range and a higher damage risk due to increased likeliness of direct beam impacts and high voltages at high charge mode. Special attention must be given to machine protection for all subsequent components.

\subsection{Aperture reduction}
\begin{figure}
	\includegraphics[width=1\columnwidth]{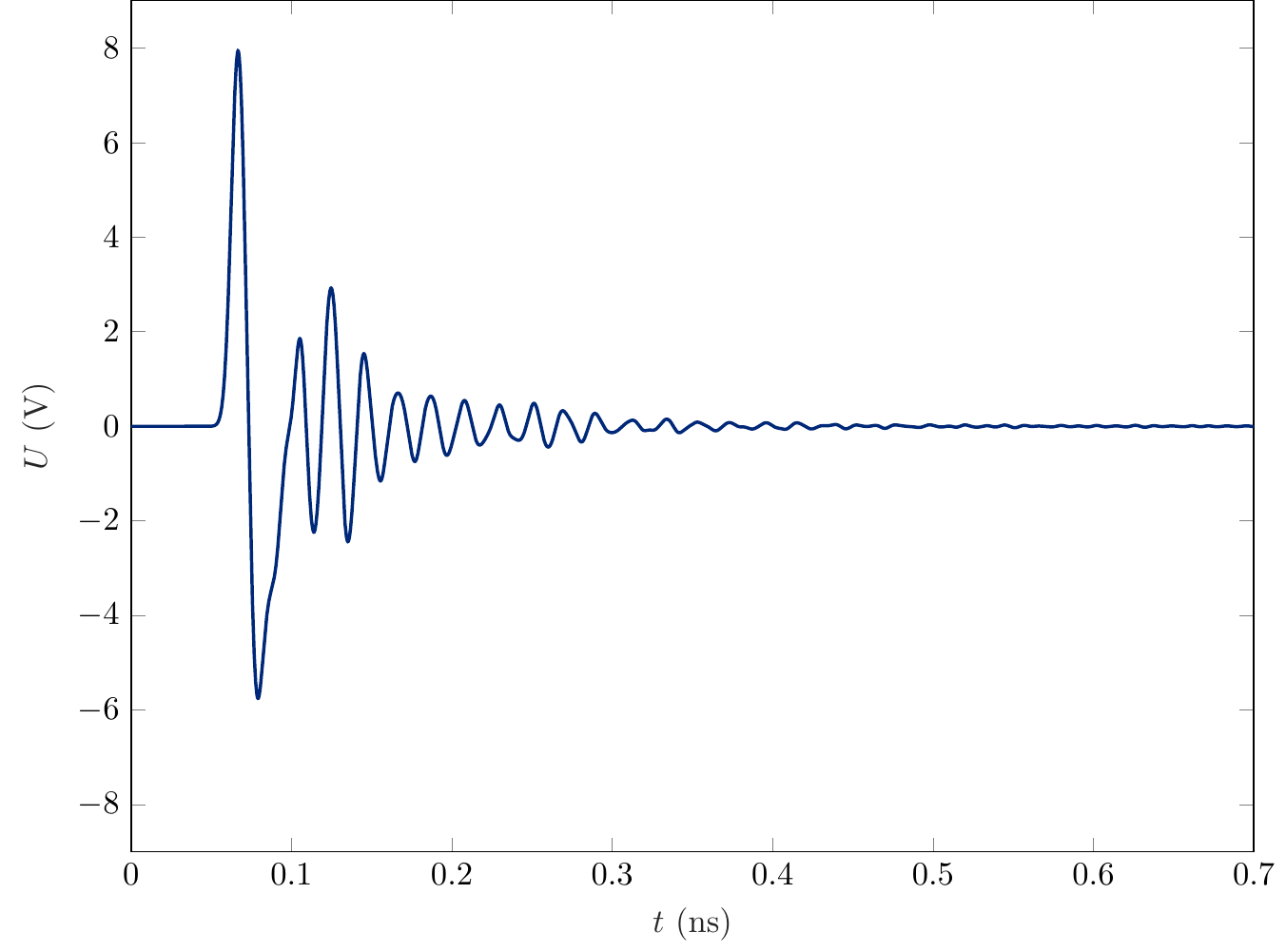}\\
	\vspace{5mm}
	\includegraphics[width=1\columnwidth]{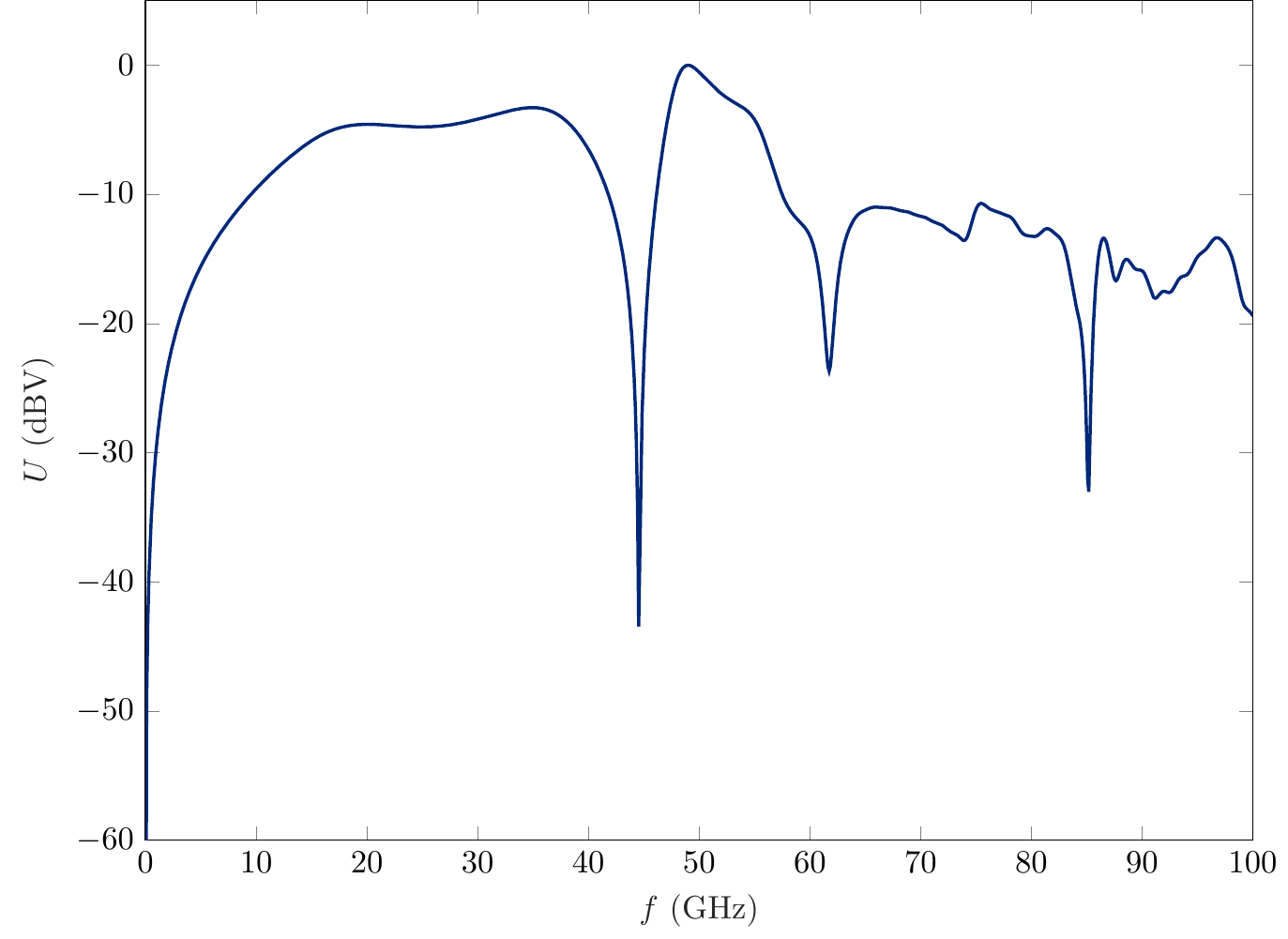}\\
	\caption{Simulated signal in time domain (top) and its normalized spectrum (bottom) taken at the end of the vacuum feedthrough of a single pickup with \SI{10}{\milli\meter} minimum distance, adapted from \cite{Scheible-2020}.\label{fig:Signal_smallG1}}
\end{figure}

A straightforward option for improvement is the sole reduction of the beam pipe aperture. A diameter of \SI{10}{\milli\meter} is possible in combination with the smaller first-generation cone-shaped pickup. The second-generation pickups cannot be installed in a four-pickup configuration with this diameter due to their dimensions, as the cut-outs would overlap. A simulation with the wakefield solver of CST PARTICLE STUDIO\texttrademark \ yields a slope of \SI{1746}{\milli\volt\per\pico\second} with \SI{13.7}{\volt} peak-to-peak voltage and a dynamical range of \SI{12.4}{\pico\second} at the nominal \SI{20}{\pico\coulomb}. The ringing of the signal pictured in Fig. \ref{fig:Signal_smallG1} initially is significantly higher but decreases rapidly during the first \SI{0.3}{\nano\second}.

\subsection{90-GHz cone shaped pickup}
We proposed a concept with pickups scaled to support up to \SI{90}{\giga\hertz} in 2019 with the dimensions specified in Table \ref{tab:PickupSpecs} \cite{Penirschke-2019}. This pickup was simulated afterwards in a BAM like setup, with four identical pickups equally distributed around a pipe section, omitting the proposed signal combination. In addition, the former bunch parameters have been initially used. These are a super relativistic (\(v = c_0\)) Gaussian bunch with \(\sigma_{\mathrm{z}}=\SI{1}{\milli\meter}\) and a charge of \SI{20}{\pico\coulomb}. Therefore, the voltage can be compared to the well-described state-of-the-art pickups according to Angelovski et al. \cite{Angelovski-2013, Angelovski-2015}.

\begin{table}
\caption{Specifications of the \SI{90}{\giga\hertz}-pickup \cite{Penirschke-2019}, the modified pickup (2\textsuperscript{nd}~Gen.) \cite{Angelovski-2015} and the original (1\textsuperscript{st}~Gen.) \cite{Angelovski-2012}.\label{tab:PickupSpecs}}
\begin{ruledtabular}
\begin{tabular}{l c c c}
	& {Draft'19} & {2\textsuperscript{nd}~Gen.} & {1\textsuperscript{st}~Gen.} \\ \hline
	Cut-out dia. (\si{mm})     		& \tablenum{1.00}	& \tablenum{1.62}	& \tablenum{1.62}	\\
	Tapered cut-out dia. (\si{mm}) 	& \tablenum{2.26}	& \tablenum{13.6}	& \tablenum{5.60}	\\
	Cone dia. (\si{mm})         		& \tablenum{0.45}	& \tablenum{0.70}	& \tablenum{0.70}	\\
	Tapered cone dia. (\si{mm})   	& \tablenum{1.02}	& \tablenum{6}		& \tablenum{2.42}	\\
	Cone height  (\si{mm})   			& {\textbackslash\footnote[1]{Not specified in the publication}} & \tablenum{6}& \tablenum{6}\\
	Protrusion  (\si{mm})       		& {\textbackslash\footnotemark}	& \tablenum{1} 		& \tablenum{0}		\\
	Relative permittivity 			& \tablenum{3.75}	& \tablenum{4.1} 	& \tablenum{4.1} 	\\
	Line impedance  (\si{\ohm})       & \tablenum{50.0}		& \tablenum{50.0} 		& \tablenum{50.0}		\\
\end{tabular}
\end{ruledtabular}
\end{table}

\begin{figure}
	\includegraphics[width=1\columnwidth]{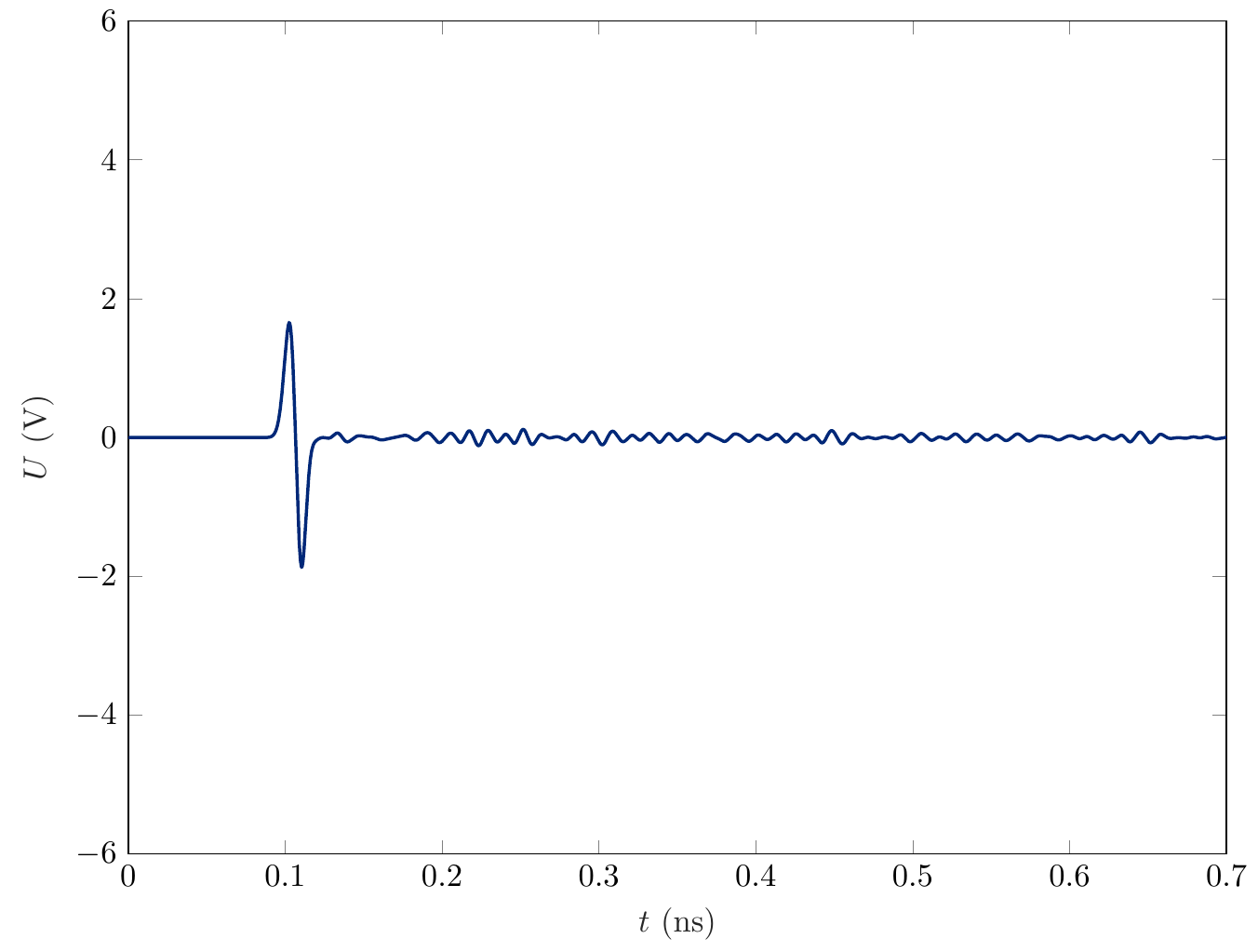}\\
	\vspace{5mm}
	\includegraphics[width=1\columnwidth]{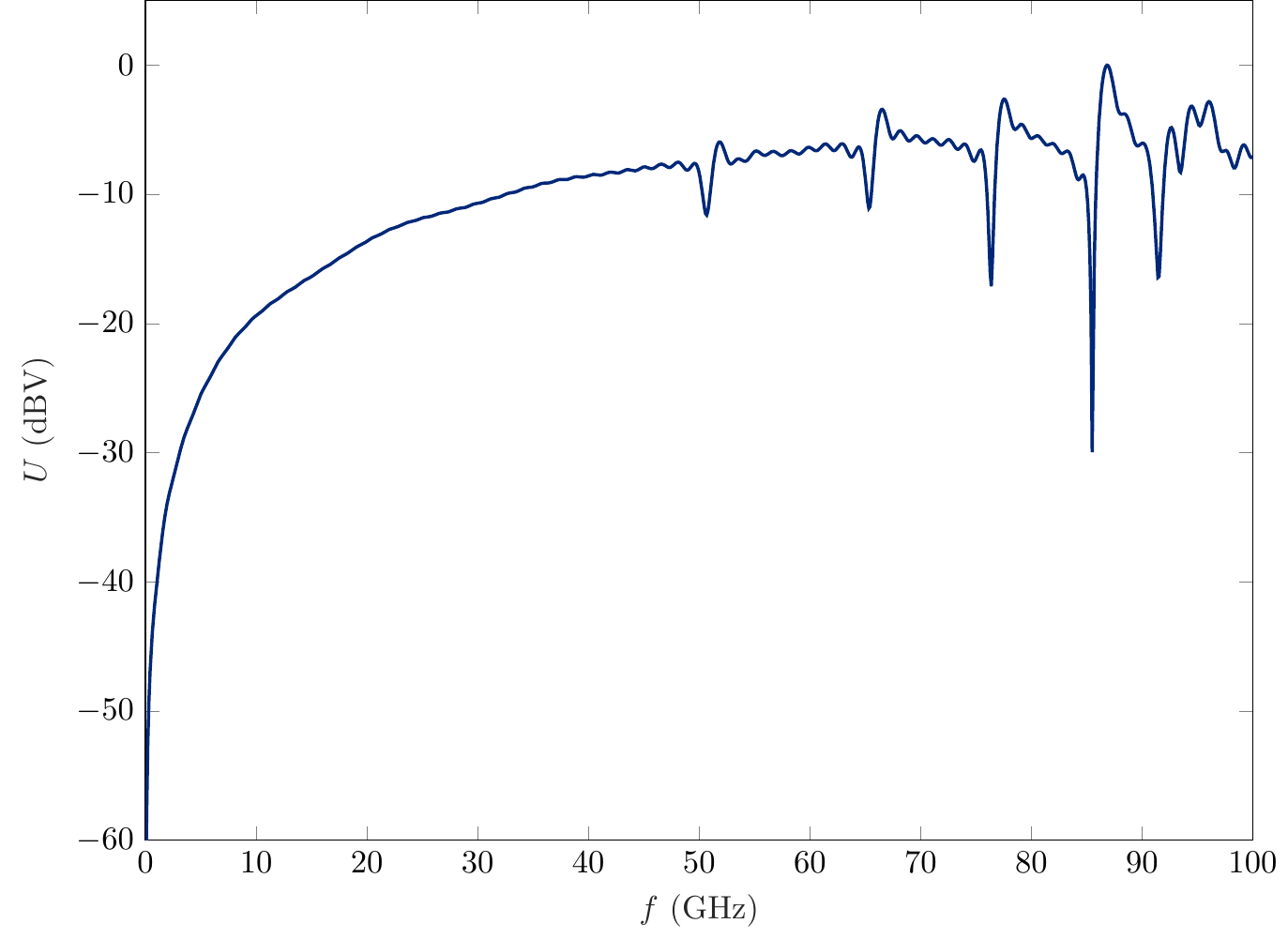}\\
	\vspace{5mm}
	\includegraphics[width=0.65\columnwidth]{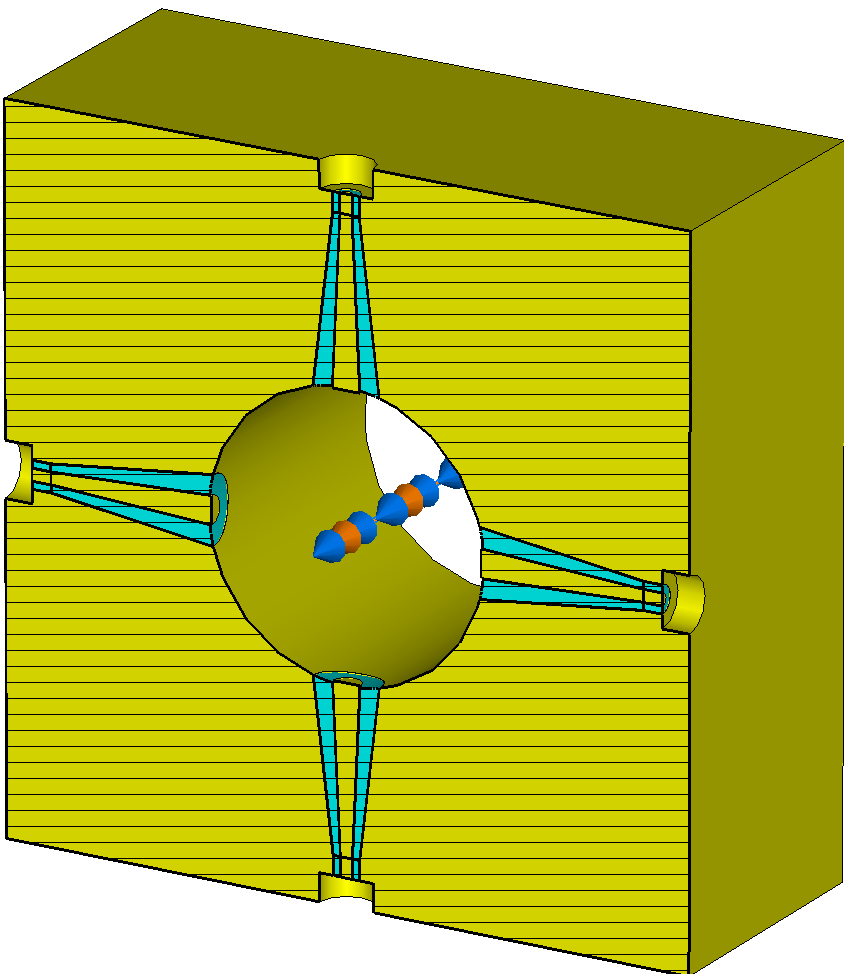}
	\caption{The signal in time domain (top) and its normalized spectrum (center) taken at the end of the vacuum feedthrough of a single pickup as well as the simulation model with \SI{90}{\giga\hertz} cone-shaped pickups (bottom), adapted from \cite{Scheible-2020}.\label{fig:IBIC19-PU}}
\end{figure}

The bipolar signal pictured in Fig. \ref{fig:IBIC19-PU} has a peak-to-peak voltage of \SI{3.52}{\volt}. The peaks are separated by \SI{8.01}{\pico\second}, more than doubling the slope to \SI{722.4}{\milli\volt\per\pico\second}. Though the pipe radius was decreased nearly by a factor of four, the gain is only by \num{2.4}. The increased bandwidth leads to a reduced rise time of \SI{7.9}{\pico\second} but cannot compensate for the missing protrusion and the smaller active area. Thus, the pickup cannot outperform the 1\textsuperscript{st} generation, but a smaller peak-to-peak voltage gives advantages in machine protection.

Reducing the bunch charge to \SI{1}{\pico\coulomb} accordingly gives a slope of \SI{36.2}{\milli\volt\per\pico\second}, which undershoots the minimum target by about a factor of \num{4}. For a further increase the combination of more than two signals is planned. By the combination of \num{8} pickups, without any phase shift and \SI{3}{\decibel} attenuation at each stage, a factor of \num{2.8} might be possible. Eight pickups has been determined as the limit, because \SI{400}{\ohm} pickups would be necessary due to the impedance change at a T junction type combiner, to have a \SI{50}{\ohm} connection to the EOM. This is well above the vacuum impedance and requires tiny components. A less radical option is the combination of \num{4} pickups for a potential improvement of a factor of \num{2}.

Compared to both preceding pickup generations, the voltage is approximately proportional to the radius of the circular pickup surface. This may indicate a bunch-pickup ratio where only the pickup width, but not its length is relevant.

\begin{figure}
	\includegraphics[width=1\columnwidth]{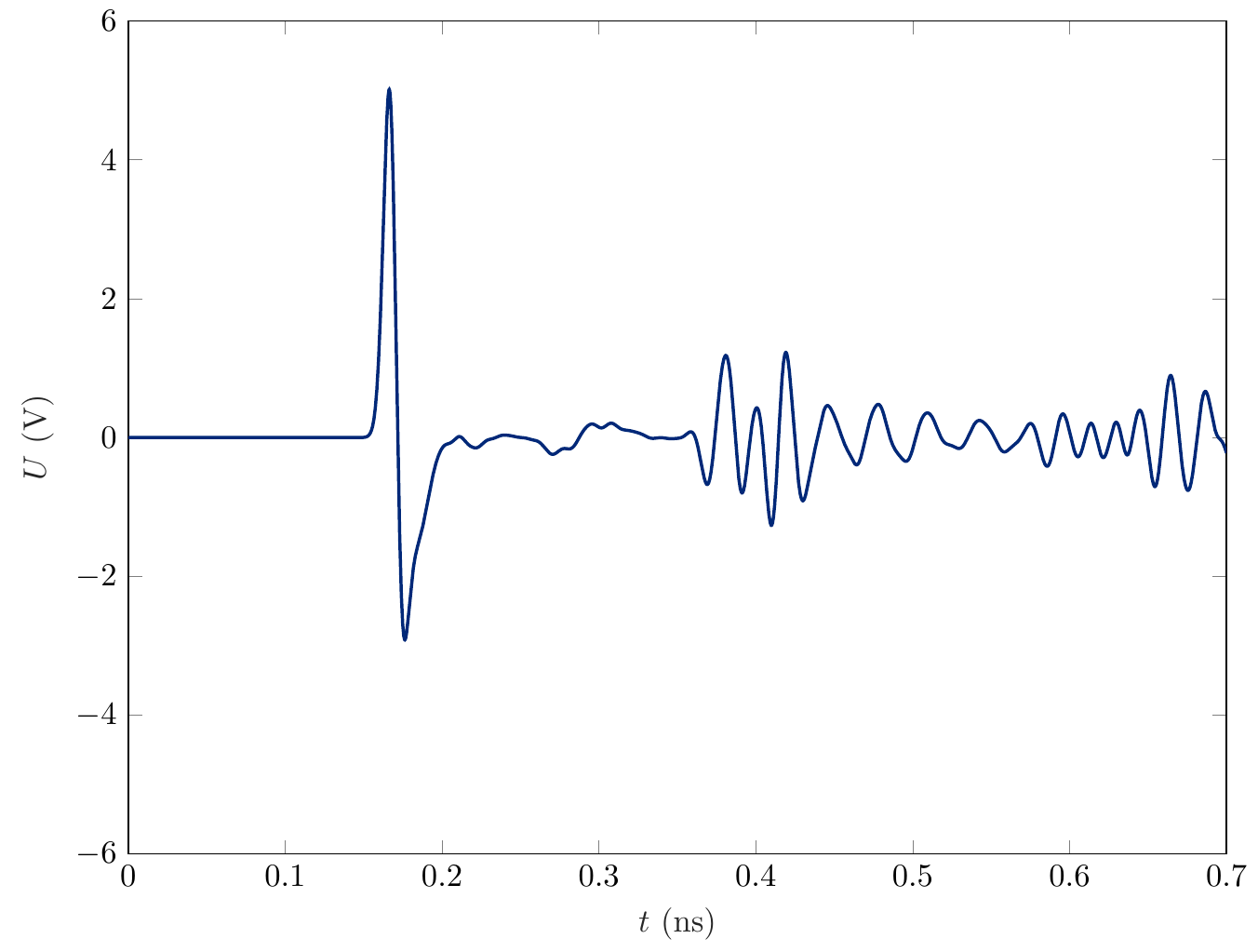}\\
	\vspace{5mm}
	\includegraphics[width=1\columnwidth]{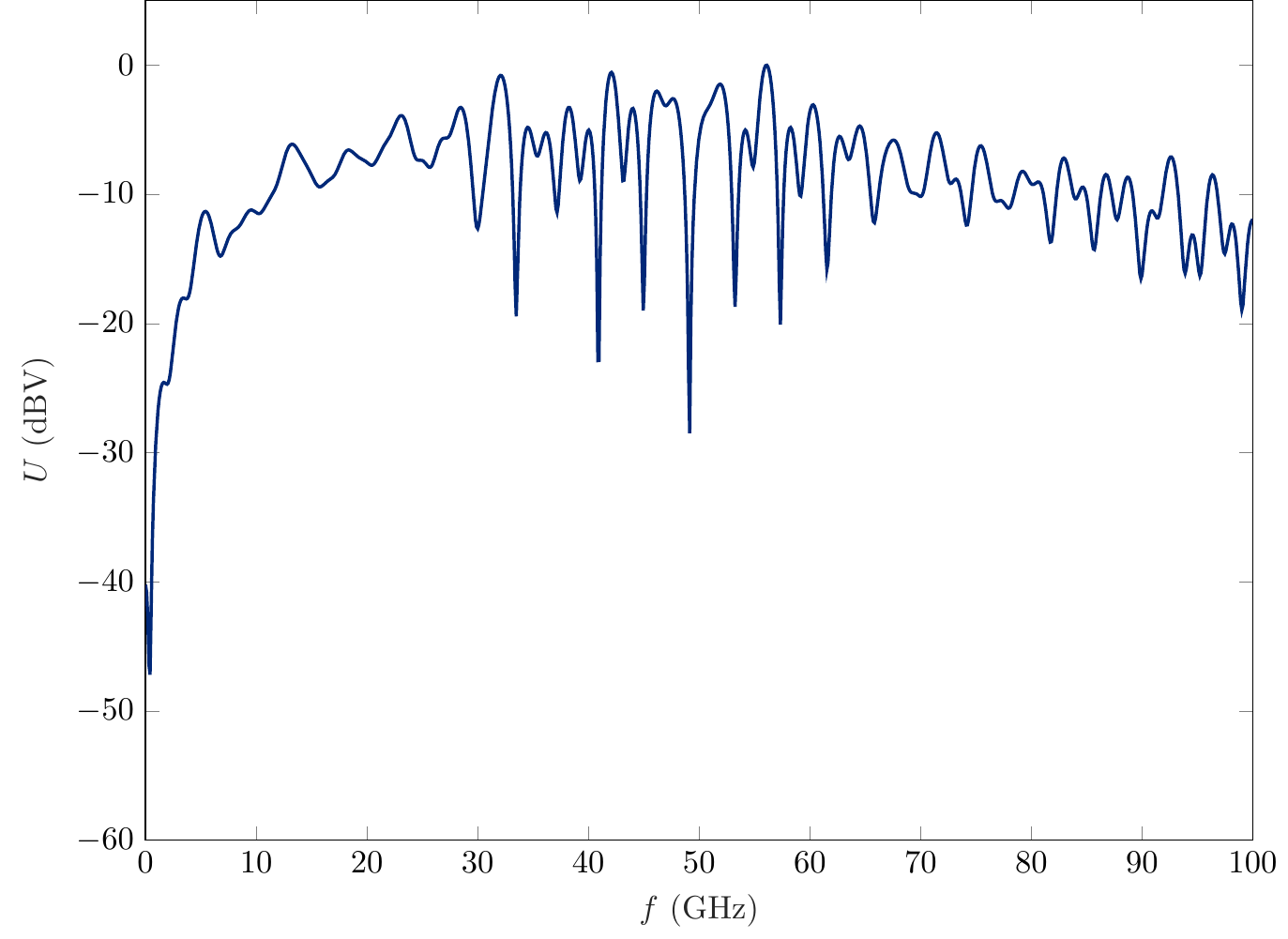}\\
	\vspace{5mm}
	\includegraphics[width=0.82\columnwidth]{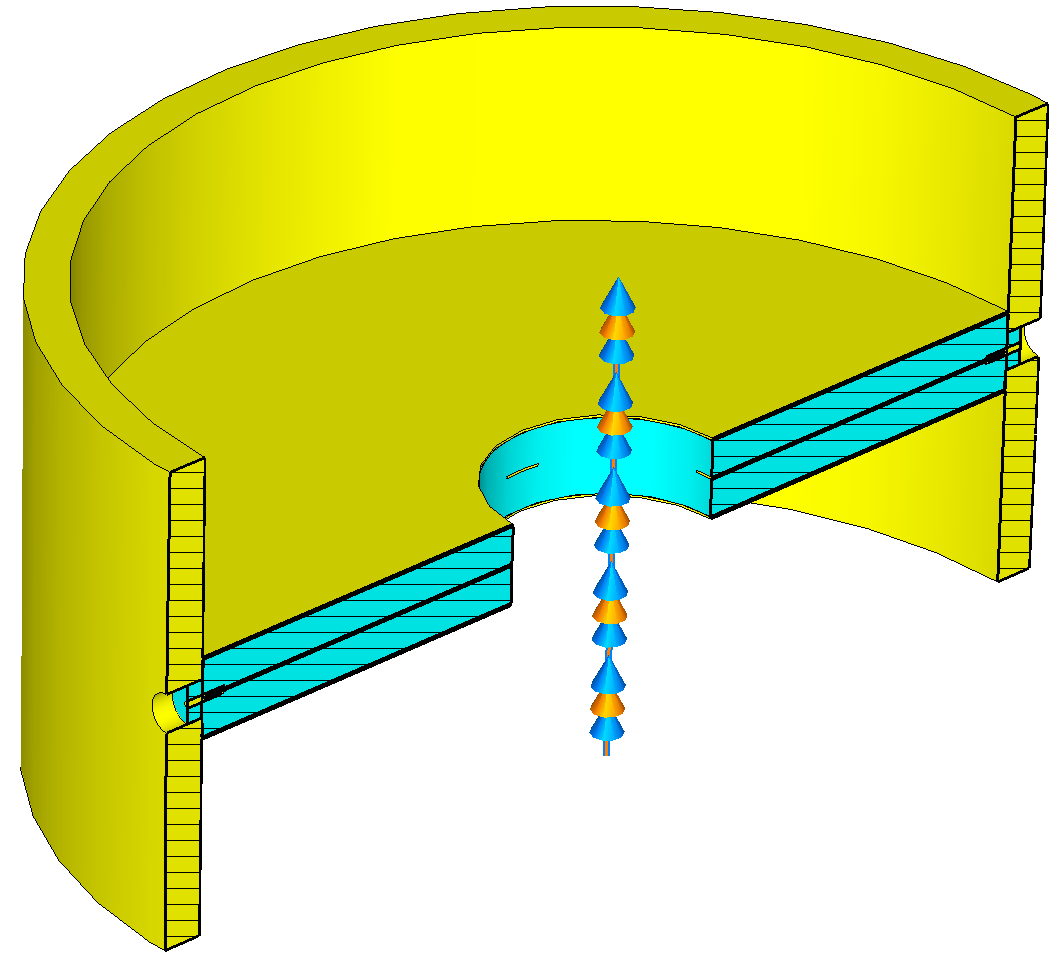}
	\caption{The signal in time domain (top) and its normalized spectrum (center) taken at the end of the vacuum feedthrough of a single pickup as well as the simulation model with \SI{50}{\ohm} stripline pickups (bottom), adapted from \cite{Scheible-2020}.\label{fig:SL-PU}}
\end{figure}

\subsection{Printed circuit board BAM}
For ultra-short bunches the transition to a short rectangular pickup on a printed circuit board (PCB) with a trace thickness still larger than the bunch length appears possible. This concept possibly allows for a \SI{100}{\giga\hertz} pickup without the drawback of smaller dimensions. Further benefits are the possibility to use well-known components with precise production methods and well-described materials. The transmission lines (TL) and combination network may be realized on the PCB reducing the RF path, which is specifically important to prevent dispersion effects in broadband quasi-TEM TLs. A microstrip is favorable for its width, but entails dispersion and is less shielded. Therefore, it is planned to use a microstrip (MS) for coupling to the field and a stripline (SL) for the combination network.

A preliminary simulation, shown in Fig. \ref{fig:SL-PU}, of a PCB based pickup was done with a \SI{1.55}{\milli\meter} wide and about \SI{15}{\milli\meter} long \SI{50}{\ohm} SL in a \(\epsilon_\mathrm{r} = \num{4.03}\) substrate disc with \SI{10}{\milli\meter} aperture inside. For a \SI{20}{\pico\coulomb} bunch the simulation returns a slope of about \SI{1270}{\milli\volt\per\pico\second}. The SL pickup is exceeding the current as well as the \SI{90}{\giga\hertz} cone-shaped pickups, but does not achieve the performance of a generation 1 pickup of equal aperture. Furthermore, crosstalk and reflections at the vacuum feedthrough as well as the open pickup end generate delayed but significant ringing.

\section{Conclusion}
A high bandwidth cone-shaped pickup with \SI{10}{\milli\meter} aperture leads to a significant improvement by reduction of the distance and increase of the bandwidth. If a maximum voltage is of no concern, a configuration of 1\textsubscript{st} generation pickups in a \SI{10}{\milli\meter} beam pipe is a simple solution estimated sufficient for bunch charges down to \SI{4}{\pico\coulomb}. In case of ultra-short bunches a PCB-type BAM may be suited to support \SI{100}{\giga\hertz} without the drawback of reduced dimensions. With the current design restrictions, a signal combination is inevitable. Further studies of an integrated combination network are required to reduce signal reflections and losses. Furthermore, it is necessary to investigate the properties of PCB boards regarding vacuum suitability and radiation hardness. Specifically, potential damages caused by beam incidence need to be assessed.

\begin{acknowledgments}
This work is supported by the German Federal Ministry of Education and Research (BMBF) under contract no. 05K19RO1.

The contribution is a version of \cite{Scheible-2020}, edited for publication in the Physical Review Accelerators and Beams (PRAB) Special Edition for the 9th International Beam Instrumentation Conference, IBIC 2020.
\end{acknowledgments}

\bibliography{PRAB_PAPER.bib}

\providecommand{\noopsort}[1]{}\providecommand{\singleletter}[1]{#1}%
\begin{thebibliography}{28}%
\makeatletter
\providecommand \@ifxundefined [1]{%
 \@ifx{#1\undefined}
}%
\providecommand \@ifnum [1]{%
 \ifnum #1\expandafter \@firstoftwo
 \else \expandafter \@secondoftwo
 \fi
}%
\providecommand \@ifx [1]{%
 \ifx #1\expandafter \@firstoftwo
 \else \expandafter \@secondoftwo
 \fi
}%
\providecommand \natexlab [1]{#1}%
\providecommand \enquote  [1]{``#1''}%
\providecommand \bibnamefont  [1]{#1}%
\providecommand \bibfnamefont [1]{#1}%
\providecommand \citenamefont [1]{#1}%
\providecommand \href@noop [0]{\@secondoftwo}%
\providecommand \href [0]{\begingroup \@sanitize@url \@href}%
\providecommand \@href[1]{\@@startlink{#1}\@@href}%
\providecommand \@@href[1]{\endgroup#1\@@endlink}%
\providecommand \@sanitize@url [0]{\catcode `\\12\catcode `\$12\catcode
  `\&12\catcode `\#12\catcode `\^12\catcode `\_12\catcode `\%12\relax}%
\providecommand \@@startlink[1]{}%
\providecommand \@@endlink[0]{}%
\providecommand \url  [0]{\begingroup\@sanitize@url \@url }%
\providecommand \@url [1]{\endgroup\@href {#1}{\urlprefix }}%
\providecommand \urlprefix  [0]{URL }%
\providecommand \Eprint [0]{\href }%
\providecommand \doibase [0]{https://doi.org/}%
\providecommand \selectlanguage [0]{\@gobble}%
\providecommand \bibinfo  [0]{\@secondoftwo}%
\providecommand \bibfield  [0]{\@secondoftwo}%
\providecommand \translation [1]{[#1]}%
\providecommand \BibitemOpen [0]{}%
\providecommand \bibitemStop [0]{}%
\providecommand \bibitemNoStop [0]{.\EOS\space}%
\providecommand \EOS [0]{\spacefactor3000\relax}%
\providecommand \BibitemShut  [1]{\csname bibitem#1\endcsname}%
\let\auto@bib@innerbib\@empty
\bibitem [{\citenamefont {Jaeschke}\ \emph {et~al.}(2020)\citenamefont
  {Jaeschke}, \citenamefont {Khan}, \citenamefont {Schneider},\ and\
  \citenamefont {Hastings}}]{jaeschke-2020}%
  \BibitemOpen
  \bibfield  {author} {\bibinfo {author} {\bibfnamefont {E.~J.}\ \bibnamefont
  {Jaeschke}}, \bibinfo {author} {\bibfnamefont {S.}~\bibnamefont {Khan}},
  \bibinfo {author} {\bibfnamefont {J.~R.}\ \bibnamefont {Schneider}},\ and\
  \bibinfo {author} {\bibfnamefont {J.~B.}\ \bibnamefont {Hastings}},\ }\href
  {https://doi.org/\url{10.1007/978-3-030-23201-6}} {\emph {\bibinfo {title}
  {{Synchrotron Light Sources and Free-Electron Lasers: Accelerator Physics,
  Instrumentation and Science Applications}}}},\ \bibinfo {edition} {2nd}\ ed.\
  (\bibinfo  {publisher} {{Springer International Publishing}},\ \bibinfo
  {address} {Cham},\ \bibinfo {year} {2020})\BibitemShut {NoStop}%
\bibitem [{\citenamefont {Seddon}\ \emph {et~al.}(2017)\citenamefont {Seddon}
  \emph {et~al.}}]{Seddon-2017}%
  \BibitemOpen
  \bibfield  {author} {\bibinfo {author} {\bibfnamefont {E.~A.}\ \bibnamefont
  {Seddon}} \emph {et~al.},\ }\bibfield  {title} {\bibinfo {title}
  {{Short-wavelength free-electron laser sources and science: a review}},\
  }\href {https://doi.org/\url{10.1088/1361-6633/aa7cca}} {\bibfield  {journal}
  {\bibinfo  {journal} {{Reports on progress in physics. Physical Society
  (Great Britain)}}\ }\textbf {\bibinfo {volume} {80}},\ \bibinfo {pages}
  {115901} (\bibinfo {year} {2017})}\BibitemShut {NoStop}%
\bibitem [{\citenamefont {G{\"u}nther}\ \emph {et~al.}(2011)\citenamefont
  {G{\"u}nther} \emph {et~al.}}]{Gunther-2011}%
  \BibitemOpen
  \bibfield  {author} {\bibinfo {author} {\bibfnamefont {C.~M.}\ \bibnamefont
  {G{\"u}nther}} \emph {et~al.},\ }\bibfield  {title} {\bibinfo {title}
  {{Sequential femtosecond X-ray imaging}},\ }\href
  {https://doi.org/\url{10.1038/nphoton.2010.287}} {\bibfield  {journal}
  {\bibinfo  {journal} {{Nature Photonics}}\ }\textbf {\bibinfo {volume} {5}},\
  \bibinfo {pages} {99} (\bibinfo {year} {2011})}\BibitemShut {NoStop}%
\bibitem [{\citenamefont {Lu}\ \emph {et~al.}(2018)\citenamefont {Lu} \emph
  {et~al.}}]{Lu-2018}%
  \BibitemOpen
  \bibfield  {author} {\bibinfo {author} {\bibfnamefont {W.}~\bibnamefont {Lu}}
  \emph {et~al.},\ }\bibfield  {title} {\bibinfo {title} {{Development of a
  hard X-ray split-and-delay line and performance simulations for two-color
  pump-probe experiments at the European XFEL}},\ }\href
  {https://doi.org/\url{10.1063/1.5027071}} {\bibfield  {journal} {\bibinfo
  {journal} {{The Review of scientific instruments}}\ }\textbf {\bibinfo
  {volume} {89}},\ \bibinfo {pages} {063121} (\bibinfo {year}
  {2018})}\BibitemShut {NoStop}%
\bibitem [{\citenamefont {Rosenzweig}\ \emph {et~al.}(2008)\citenamefont
  {Rosenzweig} \emph {et~al.}}]{Rosenzweig-2008}%
  \BibitemOpen
  \bibfield  {author} {\bibinfo {author} {\bibfnamefont {J.~B.}\ \bibnamefont
  {Rosenzweig}} \emph {et~al.},\ }\bibfield  {title} {\bibinfo {title}
  {{Generation of ultra-short, high brightness electron beams for single-spike
  SASE FEL operation}},\ }\href
  {https://doi.org/\url{10.1016/j.nima.2008.04.083}} {\bibfield  {journal}
  {\bibinfo  {journal} {{Nuclear Instruments and Methods in Physics Research
  Section A: Accelerators, Spectrometers, Detectors and Associated Equipment}}\
  }\textbf {\bibinfo {volume} {593}},\ \bibinfo {pages} {39} (\bibinfo {year}
  {2008})}\BibitemShut {NoStop}%
\bibitem [{\citenamefont {Reiche}\ \emph {et~al.}(2008)\citenamefont {Reiche},
  \citenamefont {Musumeci}, \citenamefont {Pellegrini},\ and\ \citenamefont
  {Rosenzweig}}]{Reiche-2008}%
  \BibitemOpen
  \bibfield  {author} {\bibinfo {author} {\bibfnamefont {S.}~\bibnamefont
  {Reiche}}, \bibinfo {author} {\bibfnamefont {P.}~\bibnamefont {Musumeci}},
  \bibinfo {author} {\bibfnamefont {C.}~\bibnamefont {Pellegrini}},\ and\
  \bibinfo {author} {\bibfnamefont {J.~B.}\ \bibnamefont {Rosenzweig}},\
  }\bibfield  {title} {\bibinfo {title} {{Development of ultra-short pulse,
  single coherent spike for SASE X-ray FELs}},\ }\href
  {https://doi.org/\url{10.1016/j.nima.2008.04.061}} {\bibfield  {journal}
  {\bibinfo  {journal} {{Nuclear Instruments and Methods in Physics Research
  Section A: Accelerators, Spectrometers, Detectors and Associated Equipment}}\
  }\textbf {\bibinfo {volume} {593}},\ \bibinfo {pages} {45} (\bibinfo {year}
  {2008})}\BibitemShut {NoStop}%
\bibitem [{\citenamefont {Decking}\ and\ \citenamefont
  {Limberg}(2013)}]{Decking-2013}%
  \BibitemOpen
  \bibfield  {author} {\bibinfo {author} {\bibfnamefont {W.}~\bibnamefont
  {Decking}}\ and\ \bibinfo {author} {\bibfnamefont {T.}~\bibnamefont
  {Limberg}},\ }\href@noop {} {\emph {\bibinfo {title} {{European XFEL Post TDR
  Description: XFEL.EU TN-2013-004-01}}}},\ \bibinfo {type} {Tech. Rep.}\
  (\bibinfo  {institution} {{European XFEL GmbH}},\ \bibinfo {address}
  {Hamburg, Germany},\ \bibinfo {year} {2013})\BibitemShut {NoStop}%
\bibitem [{\citenamefont {Tschentscher}(2011)}]{Tschentscher-2011}%
  \BibitemOpen
  \bibfield  {author} {\bibinfo {author} {\bibfnamefont {T.}~\bibnamefont
  {Tschentscher}},\ }\href {https://doi.org/\url{10.3204/XFEL.EU/TR-2011-001}}
  {\emph {\bibinfo {title} {{Layout of the X-Ray Systems at the European XFEL:
  TECHNICAL REPORT: XFEL.EU TR-2011-001}}}},\ \bibinfo {type} {Tech. Rep.}\
  (\bibinfo  {institution} {{Deutsches Elektronen-Synchrotron, DESY}},\
  \bibinfo {address} {Hamburg, Germany},\ \bibinfo {year} {2011})\BibitemShut
  {NoStop}%
\bibitem [{\citenamefont {Kot}\ \emph {et~al.}(2013)\citenamefont {Kot},
  \citenamefont {Limberg},\ and\ \citenamefont {Zagorodnov}}]{Kot-2013}%
  \BibitemOpen
  \bibfield  {author} {\bibinfo {author} {\bibfnamefont {Y.}~\bibnamefont
  {Kot}}, \bibinfo {author} {\bibfnamefont {T.}~\bibnamefont {Limberg}},\ and\
  \bibinfo {author} {\bibfnamefont {I.}~\bibnamefont {Zagorodnov}},\
  }\href@noop {} {\emph {\bibinfo {title} {{Technical report: Different charges
  in the same bunch train at the European XFEL}}}},\ \bibinfo {type} {Tech.
  Rep.}\ (\bibinfo  {institution} {{Deutsches Elektronen-Synchrotron, DESY}},\
  \bibinfo {address} {Hamburg, Germany},\ \bibinfo {year} {2013})\ \bibinfo
  {note} {{DESY-13-215}}\BibitemShut {NoStop}%
\bibitem [{\citenamefont {Viti}\ \emph {et~al.}(2017)\citenamefont {Viti},
  \citenamefont {Czwalinna}, \citenamefont {Dinter}, \citenamefont {Gerth},
  \citenamefont {Przygoda}, \citenamefont {Rybaniec},\ and\ \citenamefont
  {Schlarb}}]{Viti-2017}%
  \BibitemOpen
  \bibfield  {author} {\bibinfo {author} {\bibfnamefont {M.}~\bibnamefont
  {Viti}}, \bibinfo {author} {\bibfnamefont {M.}~\bibnamefont {Czwalinna}},
  \bibinfo {author} {\bibfnamefont {H.}~\bibnamefont {Dinter}}, \bibinfo
  {author} {\bibfnamefont {C.}~\bibnamefont {Gerth}}, \bibinfo {author}
  {\bibfnamefont {K.}~\bibnamefont {Przygoda}}, \bibinfo {author}
  {\bibfnamefont {R.}~\bibnamefont {Rybaniec}},\ and\ \bibinfo {author}
  {\bibfnamefont {H.}~\bibnamefont {Schlarb}},\ }\bibfield  {title} {\bibinfo
  {title} {{Recent Upgrades of the Bunch Arrival Time Monitors at FLASH and
  European XFEL}},\ }in\ \href
  {https://doi.org/https://doi.org/10.18429/JACoW-IPAC2017-MOPIK072} {\emph
  {\bibinfo {booktitle} {Proc. of 8th International Particle Accelerator
  Conference, IPAC'17}}}\ (\bibinfo  {publisher} {JACoW},\ \bibinfo {address}
  {Geneva, Switzerland},\ \bibinfo {year} {2017})\ pp.\ \bibinfo {pages}
  {695--698}\BibitemShut {NoStop}%
\bibitem [{\citenamefont {L{\"o}hl}\ \emph {et~al.}(2010)\citenamefont
  {L{\"o}hl} \emph {et~al.}}]{Lohl-2010}%
  \BibitemOpen
  \bibfield  {author} {\bibinfo {author} {\bibfnamefont {F.}~\bibnamefont
  {L{\"o}hl}} \emph {et~al.},\ }\bibfield  {title} {\bibinfo {title} {{Electron
  bunch timing with femtosecond precision in a superconducting free-electron
  laser}},\ }\href {https://doi.org/\url{10.1103/PhysRevLett.104.144801}}
  {\bibfield  {journal} {\bibinfo  {journal} {{Physical review letters}}\
  }\textbf {\bibinfo {volume} {104}},\ \bibinfo {pages} {144801} (\bibinfo
  {year} {2010})}\BibitemShut {NoStop}%
\bibitem [{\citenamefont {Angelovski}\ \emph {et~al.}(2012)\citenamefont
  {Angelovski}, \citenamefont {Kuhl}, \citenamefont {Hansli}, \citenamefont
  {Penirschke}, \citenamefont {Schnepp}, \citenamefont {Bousonville},
  \citenamefont {Schlarb}, \citenamefont {Bock}, \citenamefont {Weiland},\ and\
  \citenamefont {Jakoby}}]{Angelovski-2012}%
  \BibitemOpen
  \bibfield  {author} {\bibinfo {author} {\bibfnamefont {A.}~\bibnamefont
  {Angelovski}}, \bibinfo {author} {\bibfnamefont {A.}~\bibnamefont {Kuhl}},
  \bibinfo {author} {\bibfnamefont {M.}~\bibnamefont {Hansli}}, \bibinfo
  {author} {\bibfnamefont {A.}~\bibnamefont {Penirschke}}, \bibinfo {author}
  {\bibfnamefont {S.~M.}\ \bibnamefont {Schnepp}}, \bibinfo {author}
  {\bibfnamefont {M.}~\bibnamefont {Bousonville}}, \bibinfo {author}
  {\bibfnamefont {H.}~\bibnamefont {Schlarb}}, \bibinfo {author} {\bibfnamefont
  {M.~K.}\ \bibnamefont {Bock}}, \bibinfo {author} {\bibfnamefont
  {T.}~\bibnamefont {Weiland}},\ and\ \bibinfo {author} {\bibfnamefont
  {R.}~\bibnamefont {Jakoby}},\ }\bibfield  {title} {\bibinfo {title} {High
  bandwidth pickup design for bunch arrival-time monitors for free-electron
  laser},\ }\href {https://doi.org/10.1103/PhysRevSTAB.15.112803} {\bibfield
  {journal} {\bibinfo  {journal} {Phys. Rev. ST Accel. Beams}\ }\textbf
  {\bibinfo {volume} {15}},\ \bibinfo {pages} {112803} (\bibinfo {year}
  {2012})}\BibitemShut {NoStop}%
\bibitem [{\citenamefont {Angelovski}\ \emph {et~al.}(2013)\citenamefont
  {Angelovski}, \citenamefont {Penirschke}, \citenamefont {Jakoby},
  \citenamefont {Czwalinna}, \citenamefont {Sydlo}, \citenamefont {Schlarb},\
  and\ \citenamefont {Weiland}}]{Angelovski-2013}%
  \BibitemOpen
  \bibfield  {author} {\bibinfo {author} {\bibfnamefont {A.}~\bibnamefont
  {Angelovski}}, \bibinfo {author} {\bibfnamefont {A.}~\bibnamefont
  {Penirschke}}, \bibinfo {author} {\bibfnamefont {R.}~\bibnamefont {Jakoby}},
  \bibinfo {author} {\bibfnamefont {M.~K.}\ \bibnamefont {Czwalinna}}, \bibinfo
  {author} {\bibfnamefont {C.}~\bibnamefont {Sydlo}}, \bibinfo {author}
  {\bibfnamefont {H.}~\bibnamefont {Schlarb}},\ and\ \bibinfo {author}
  {\bibfnamefont {T.}~\bibnamefont {Weiland}},\ }\bibfield  {title} {\bibinfo
  {title} {Pickup signal improvement for high bandwidth bams for flash and
  european - xfel},\ }in\ \href {http://tubiblio.ulb.tu-darmstadt.de/79291/}
  {\emph {\bibinfo {booktitle} {Proceedings of the International Beam
  Instrumentation Conference, IBIC'13}}}\ (\bibinfo  {publisher} {JACoW},\
  \bibinfo {address} {Geneva, Switzerland},\ \bibinfo {year} {2013})\ pp.\
  \bibinfo {pages} {778--781}\BibitemShut {NoStop}%
\bibitem [{\citenamefont {Kim}\ \emph {et~al.}(2006)\citenamefont {Kim},
  \citenamefont {Burnham}, \citenamefont {Chen}, \citenamefont {Kartner},
  \citenamefont {Ilday}, \citenamefont {Ludwig}, \citenamefont {Schlarb},
  \citenamefont {Winter}, \citenamefont {Ferianis},\ and\ \citenamefont
  {Cheever}}]{Kim-2006}%
  \BibitemOpen
  \bibfield  {author} {\bibinfo {author} {\bibfnamefont {J.}~\bibnamefont
  {Kim}}, \bibinfo {author} {\bibfnamefont {J.}~\bibnamefont {Burnham}},
  \bibinfo {author} {\bibfnamefont {J.}~\bibnamefont {Chen}}, \bibinfo {author}
  {\bibfnamefont {F.~X.}\ \bibnamefont {Kartner}}, \bibinfo {author}
  {\bibfnamefont {F.~{\"O}.}\ \bibnamefont {Ilday}}, \bibinfo {author}
  {\bibfnamefont {F.}~\bibnamefont {Ludwig}}, \bibinfo {author} {\bibfnamefont
  {H.}~\bibnamefont {Schlarb}}, \bibinfo {author} {\bibfnamefont
  {A.}~\bibnamefont {Winter}}, \bibinfo {author} {\bibfnamefont
  {M.}~\bibnamefont {Ferianis}},\ and\ \bibinfo {author} {\bibfnamefont
  {D.}~\bibnamefont {Cheever}},\ }\bibfield  {title} {\bibinfo {title} {{An
  integrated femtosecond timing distribution system for XFELs}},\ }in\
  \href@noop {} {\emph {\bibinfo {booktitle} {{Proceedings of the 10th European
  Particle Accelerator Conference, EPAC'06}}}},\ \bibinfo {series and number}
  {\bibinfo {number} {10}}\ (\bibinfo  {publisher} {JACoW},\ \bibinfo {address}
  {Geneva, Switzerland},\ \bibinfo {year} {2006})\ pp.\ \bibinfo {pages}
  {2744--2746}\BibitemShut {NoStop}%
\bibitem [{\citenamefont {Lamb}\ \emph {et~al.}(2019)\citenamefont {Lamb} \emph
  {et~al.}}]{Lamb-2018}%
  \BibitemOpen
  \bibfield  {author} {\bibinfo {author} {\bibfnamefont {T.}~\bibnamefont
  {Lamb}} \emph {et~al.},\ }\bibfield  {title} {\bibinfo {title} {{Large-Scale
  Optical Synchronization System of the European XFEL with Femtosecond
  Precision: JACoW Publishing, Geneva, Switzerland}},\ }in\ \href
  {https://doi.org/\url{10.18429/JACoW-IPAC2019-THPRB018}} {\emph {\bibinfo
  {booktitle} {{Proceedings of the 10th International Particle Accelerator
  Conference, IPAC'19}}}},\ \bibinfo {organization} {10th International
  Particle Accelerator Conference, Melbourne (Australia), 19 May 2019 - 24 May
  2019}\ (\bibinfo  {publisher} {JACoW},\ \bibinfo {address} {Geneva,
  Switzerland},\ \bibinfo {year} {2019})\ pp.\ \bibinfo {pages}
  {3835--3838}\BibitemShut {NoStop}%
\bibitem [{\citenamefont {Angelovski}\ \emph {et~al.}(2015)\citenamefont
  {Angelovski} \emph {et~al.}}]{Angelovski-2015}%
  \BibitemOpen
  \bibfield  {author} {\bibinfo {author} {\bibfnamefont {A.}~\bibnamefont
  {Angelovski}} \emph {et~al.},\ }\bibfield  {title} {\bibinfo {title}
  {{Evaluation of the cone-shaped pickup performance for low charge sub-10 fs
  arrival-time measurements at free electron laser facilities}},\ }\bibfield
  {journal} {\bibinfo  {journal} {{Physical Review Special Topics -
  Accelerators and Beams}}\ }\textbf {\bibinfo {volume} {18}},\ \href
  {https://doi.org/\url{10.1103/PhysRevSTAB.18.012801}}
  {\url{10.1103/PhysRevSTAB.18.012801}} (\bibinfo {year} {2015})\BibitemShut
  {NoStop}%
\bibitem [{\citenamefont {Kuhl}\ \emph {et~al.}(2011)\citenamefont {Kuhl} \emph
  {et~al.}}]{Kuhl-2011}%
  \BibitemOpen
  \bibfield  {author} {\bibinfo {author} {\bibfnamefont {A.}~\bibnamefont
  {Kuhl}} \emph {et~al.},\ }\bibfield  {title} {\bibinfo {title} {Analysis of
  new pickup designs for the flash and xfel bunch arrival time monitor
  system},\ }in\ \href@noop {} {\emph {\bibinfo {booktitle} {Proceeding of the
  10th European Workshop on Beam Diagnostics and Instrumentation for Particle
  Accelerators, DIPAC'11}}}\ (\bibinfo  {publisher} {JACoW},\ \bibinfo
  {address} {Geneva, Switzerland},\ \bibinfo {year} {2011})\ pp.\ \bibinfo
  {pages} {125--127}\BibitemShut {NoStop}%
\bibitem [{\citenamefont {Bock}\ \emph {et~al.}(2011)\citenamefont {Bock},
  \citenamefont {Bousonville}, \citenamefont {Felber}, \citenamefont {Gessler},
  \citenamefont {Lamb}, \citenamefont {Ruzin}, \citenamefont {Schlarb},
  \citenamefont {Schmidt},\ and\ \citenamefont {Schulz}}]{Bock-2011}%
  \BibitemOpen
  \bibfield  {author} {\bibinfo {author} {\bibfnamefont {M.~K.}\ \bibnamefont
  {Bock}}, \bibinfo {author} {\bibfnamefont {M.}~\bibnamefont {Bousonville}},
  \bibinfo {author} {\bibfnamefont {M.}~\bibnamefont {Felber}}, \bibinfo
  {author} {\bibfnamefont {P.}~\bibnamefont {Gessler}}, \bibinfo {author}
  {\bibfnamefont {T.}~\bibnamefont {Lamb}}, \bibinfo {author} {\bibfnamefont
  {S.}~\bibnamefont {Ruzin}}, \bibinfo {author} {\bibfnamefont
  {H.}~\bibnamefont {Schlarb}}, \bibinfo {author} {\bibfnamefont
  {B.}~\bibnamefont {Schmidt}},\ and\ \bibinfo {author} {\bibfnamefont
  {S.}~\bibnamefont {Schulz}},\ }\bibfield  {title} {\bibinfo {title}
  {Benchmarking the performance of the present bunch arrival time monitors at
  flash},\ }in\ \href@noop {} {\emph {\bibinfo {booktitle} {Proceeding of the
  10th European Workshop on Beam Diagnostics and Instrumentation for Particle
  Accelerators, DIPAC'11}}}\ (\bibinfo  {publisher} {JACoW},\ \bibinfo
  {address} {Geneva, Switzerland},\ \bibinfo {year} {2011})\ pp.\ \bibinfo
  {pages} {365--367}\BibitemShut {NoStop}%
\bibitem [{\citenamefont {{Yin}}\ \emph {et~al.}(1995)\citenamefont {{Yin}},
  \citenamefont {{Schulte}},\ and\ \citenamefont {{Ekel\"of}}}]{Yin-1995}%
  \BibitemOpen
  \bibfield  {author} {\bibinfo {author} {\bibfnamefont {Y.}~\bibnamefont
  {{Yin}}}, \bibinfo {author} {\bibfnamefont {E.}~\bibnamefont {{Schulte}}},\
  and\ \bibinfo {author} {\bibfnamefont {T.}~\bibnamefont {{Ekel\"of}}},\
  }\bibfield  {title} {\bibinfo {title} {{Recovery of CTF beam signals from a
  strong wakefield background}},\ }in\ \href
  {https://doi.org/\url{10.1109/PAC.1995.505638}} {\emph {\bibinfo {booktitle}
  {Proc. 1995 IEEE Particle Accelerator Conference, PAC'95}}},\ Vol.~\bibinfo
  {volume} {4}\ (\bibinfo  {publisher} {JACoW},\ \bibinfo {address} {Geneva,
  Switzerland},\ \bibinfo {year} {1995})\ pp.\ \bibinfo {pages}
  {2622--2624}\BibitemShut {NoStop}%
\bibitem [{\citenamefont {Angelovski}\ \emph {et~al.}(2011)\citenamefont
  {Angelovski} \emph {et~al.}}]{Angelovski-2011}%
  \BibitemOpen
  \bibfield  {author} {\bibinfo {author} {\bibfnamefont {A.}~\bibnamefont
  {Angelovski}} \emph {et~al.},\ }\bibfield  {title} {\bibinfo {title} {Pickup
  design for a high resolution bunch arrival time monitor for flash and xfel},\
  }in\ \href@noop {} {\emph {\bibinfo {booktitle} {Proceeding of the 10th
  European Workshop on Beam Diagnostics and Instrumentation for Particle
  Accelerators, DIPAC'11}}}\ (\bibinfo  {publisher} {JACoW},\ \bibinfo
  {address} {Geneva, Switzerland},\ \bibinfo {year} {2011})\ pp.\ \bibinfo
  {pages} {122--124}\BibitemShut {NoStop}%
\bibitem [{\citenamefont {Schulz}\ \emph {et~al.}(2019)\citenamefont {Schulz}
  \emph {et~al.}}]{Schulz-2019}%
  \BibitemOpen
  \bibfield  {author} {\bibinfo {author} {\bibfnamefont {S.}~\bibnamefont
  {Schulz}} \emph {et~al.},\ }\bibfield  {title} {\bibinfo {title}
  {{Few-Femtosecond Facility-Wide Synchronization of the European XFEL}},\ }in\
  \href {https://doi.org/\url{doi:10.18429/JACoW-FEL2019-WEB04}} {\emph
  {\bibinfo {booktitle} {{Proceeding of the 39th International Free Electron
  Laser Conference, FEL'19}}}}\ (\bibinfo  {publisher} {JACoW},\ \bibinfo
  {address} {Geneva, Switzerland},\ \bibinfo {year} {2019})\ pp.\ \bibinfo
  {pages} {318--321}\BibitemShut {NoStop}%
\bibitem [{\citenamefont {Smith}(1996)}]{Smith-1996}%
  \BibitemOpen
  \bibfield  {author} {\bibinfo {author} {\bibfnamefont {S.~R.}\ \bibnamefont
  {Smith}},\ }\bibfield  {title} {\bibinfo {title} {{Beam Position Monitor
  Engineering}},\ }in\ \href {https://doi.org/\url{10.1063/1.52306}} {\emph
  {\bibinfo {booktitle} {{Proceedings of the 7th Beam Instrumentation Workhop,
  BIW'96}}}},\ \bibinfo {series and number} {\bibinfo {series} {Beam
  Instrumentation Workhop}\ No.~\bibinfo {number} {7}}\ (\bibinfo  {publisher}
  {AIP},\ \bibinfo {year} {1996})\ pp.\ \bibinfo {pages} {50--65}\BibitemShut
  {NoStop}%
\bibitem [{\citenamefont {Shafer}(1990)}]{Shafer-1990}%
  \BibitemOpen
  \bibfield  {author} {\bibinfo {author} {\bibfnamefont {R.~E.}\ \bibnamefont
  {Shafer}},\ }\bibfield  {title} {\bibinfo {title} {Beam position
  monitoring},\ }in\ \href@noop {} {\emph {\bibinfo {booktitle} {AIP conference
  proceedings}}},\ Vol.\ \bibinfo {volume} {212}\ (\bibinfo {organization}
  {American Institute of Physics},\ \bibinfo {year} {1990})\ p.\ \bibinfo
  {pages} {601–636}\BibitemShut {NoStop}%
\bibitem [{\citenamefont {Huang}\ \emph {et~al.}(2006)\citenamefont {Huang},
  \citenamefont {Yu}, \citenamefont {Hwang}, \citenamefont {Chun},\ and\
  \citenamefont {Kim}}]{Huang-2006}%
  \BibitemOpen
  \bibfield  {author} {\bibinfo {author} {\bibfnamefont {J.~Y.}\ \bibnamefont
  {Huang}}, \bibinfo {author} {\bibfnamefont {I.~H.}\ \bibnamefont {Yu}},
  \bibinfo {author} {\bibfnamefont {W.~H.}\ \bibnamefont {Hwang}}, \bibinfo
  {author} {\bibfnamefont {M.~H.}\ \bibnamefont {Chun}},\ and\ \bibinfo
  {author} {\bibfnamefont {S.~C.}\ \bibnamefont {Kim}},\ }\bibfield  {title}
  {\bibinfo {title} {{Analysis of pickup signals in the time domain and the
  frequency domain for the beam position monitor at the Pohang Light Source}},\
  }\href@noop {} {\bibfield  {journal} {\bibinfo  {journal} {{Journal of the
  Korean Physical Society}}\ }\textbf {\bibinfo {volume} {48}},\ \bibinfo
  {pages} {768} (\bibinfo {year} {2006})}\BibitemShut {NoStop}%
\bibitem [{\citenamefont {K{\"u}pfm{\"u}ller}(1932)}]{Kupfmuller-1932}%
  \BibitemOpen
  \bibfield  {author} {\bibinfo {author} {\bibfnamefont {K.}~\bibnamefont
  {K{\"u}pfm{\"u}ller}},\ }in\ \href
  {https://doi.org/\url{10.1007/978-3-662-36779-7}} {\emph {\bibinfo
  {booktitle} {{Einf{\"u}hrung in die theoretische Elek\-tro\-tech\-nik}}}}\
  (\bibinfo  {publisher} {{Springer Berlin Heidelberg}},\ \bibinfo {year}
  {1932})\ Chap.\ \bibinfo {chapter} {{40. Der Zusammenhang zwischen den
  Fre\-quenz\-cha\-rak\-te\-ris\-ti\-ken und den Ausgleichsvorg{\"a}ngen}},
  pp.\ \bibinfo {pages} {268--272}\BibitemShut {NoStop}%
\bibitem [{\citenamefont {Gabor}(1946)}]{Gabor-1946}%
  \BibitemOpen
  \bibfield  {author} {\bibinfo {author} {\bibfnamefont {D.}~\bibnamefont
  {Gabor}},\ }\bibfield  {title} {\bibinfo {title} {Theory of communication.
  part 1: The analysis of information},\ }\href
  {https://doi.org/\url{10.1049/ji-3-2.1946.0074}} {\bibfield  {journal}
  {\bibinfo  {journal} {Journal of the Institution of Electrical Engineers -
  Part III: Radio and Communication Engineering}\ }\textbf {\bibinfo {volume}
  {93}},\ \bibinfo {pages} {429} (\bibinfo {year} {1946})}\BibitemShut
  {NoStop}%
\bibitem [{\citenamefont {Scheible}\ \emph {et~al.}(2020)\citenamefont
  {Scheible}, \citenamefont {Penirschke}, \citenamefont {Czwalinna},
  \citenamefont {Schlarb}, \citenamefont {Ackermann},\ and\ \citenamefont
  {De~Gersem}}]{Scheible-2020}%
  \BibitemOpen
  \bibfield  {author} {\bibinfo {author} {\bibfnamefont {B.~E.~J.}\
  \bibnamefont {Scheible}}, \bibinfo {author} {\bibfnamefont {A.}~\bibnamefont
  {Penirschke}}, \bibinfo {author} {\bibfnamefont {M.~K.}\ \bibnamefont
  {Czwalinna}}, \bibinfo {author} {\bibfnamefont {H.}~\bibnamefont {Schlarb}},
  \bibinfo {author} {\bibfnamefont {W.}~\bibnamefont {Ackermann}},\ and\
  \bibinfo {author} {\bibfnamefont {H.}~\bibnamefont {De~Gersem}},\ }\bibfield
  {title} {\bibinfo {title} {{Evaluation of a Novel Pickup Concept for
  Ultra-Low Charged Short Bunches in X-Ray Free-Electron Lasers}},\ }in\ \href
  {https://doi.org/\url{10.18429/JACoW-IBIC2020-WEPP21}} {\emph {\bibinfo
  {booktitle} {Proceedings of the 9th International Beam Instrumentation
  Conference, IBIC'20}}}\ (\bibinfo  {publisher} {JACoW Publishing},\ \bibinfo
  {address} {Geneva, Switzerland},\ \bibinfo {year} {2020})\ pp.\ \bibinfo
  {pages} {145--149}\BibitemShut {NoStop}%
\bibitem [{\citenamefont {Penirschke}\ \emph {et~al.}(2019)\citenamefont
  {Penirschke}, \citenamefont {Ackermann}, \citenamefont {Czwalinna},
  \citenamefont {Kuntzsch},\ and\ \citenamefont {Schlarb}}]{Penirschke-2019}%
  \BibitemOpen
  \bibfield  {author} {\bibinfo {author} {\bibfnamefont {A.}~\bibnamefont
  {Penirschke}}, \bibinfo {author} {\bibfnamefont {W.}~\bibnamefont
  {Ackermann}}, \bibinfo {author} {\bibfnamefont {M.~K.}\ \bibnamefont
  {Czwalinna}}, \bibinfo {author} {\bibfnamefont {M.}~\bibnamefont
  {Kuntzsch}},\ and\ \bibinfo {author} {\bibfnamefont {H.}~\bibnamefont
  {Schlarb}},\ }\bibfield  {title} {\bibinfo {title} {{Concept of a Novel
  High-Bandwidth Arrival Time Monitor for Very Low Charges as a Part of the
  All-Optical Synchronization System at ELBE}},\ }in\ \href
  {https://doi.org/\url{10.18429/JACoW-IBIC2019-WEPP019}} {\emph {\bibinfo
  {booktitle} {{Proceedings of the 8th International Beam Instrumentation
  Conference, IBIC'19}}}}\ (\bibinfo  {publisher} {JACoW},\ \bibinfo {address}
  {Geneva, Switzerland},\ \bibinfo {year} {2019})\ pp.\ \bibinfo {pages}
  {553--556}\BibitemShut {NoStop}%
\end{thebibliography}%

\end{document}